\documentclass[11pt]{article}
\pdfoutput=1
\usepackage{amsmath}
\usepackage{graphicx}
\usepackage{natbib} \bibpunct{(}{)}{;}{a}{,}{,}
\usepackage{multicol}
\usepackage{sectsty}
\usepackage{url}
\usepackage{hyperref}
\usepackage{dcolumn}
\usepackage{caption}
\usepackage{subcaption}

\begin{document}
\par
\title{Quantum calcium-ion interactions with EEG}
\author{Lester Ingber \\
Lester Ingber Research \\
Ashland, Oregon 97520\\
ingber@alumni.caltech.edu [https://www.ingber.com]}

\date{}
\maketitle

\begin {abstract}
\textbf{Background}:
Previous papers have developed
a statistical mechanics of neocortical interactions (SMNI)
fit to short-term memory and EEG data.
Adaptive Simulated Annealing (ASA) has been developed to perform fits to such nonlinear
stochastic systems.
An N-dimensional path-integral algorithm for quantum systems, qPATHINT, has been developed from classical PATHINT.
Both fold short-time propagators (distributions
or wave functions) over long times.
Previous papers applied qPATHINT
to two systems, in neocortical interactions and
financial options.
\textbf{Objective}:
In this paper the quantum path-integral for Calcium ions is used to derive a
closed-form analytic solution at arbitrary time
that is used to calculate interactions with classical-physics SMNI interactions among scales.
Using fits of this SMNI model to EEG data, including these effects,
will help determine if this is a reasonable approach.
\textbf{Method}:
Methods of mathematical-physics for optimization and for path
integrals in classical and quantum spaces are used for this
project.
Studies using
supercomputer resources tested various dimensions for their scaling limits.
In this paper the quantum path-integral is used to derive a
closed-form analytic solution at arbitrary time
that is used to calculate interactions with classical-physics SMNI interactions among scales.
\textbf{Results}:
The mathematical-physics and computer parts of the study are successful,
in that there is modest improvement of cost/objective functions used to fit
EEG data using these models.
\textbf{Conclusion}:
This project points to directions for more detailed calculations using
more EEG data
and qPATHINT at each time slice to propagate quantum calcium waves, synchronized with PATHINT propagation of classical SMNI.
\end {abstract}

\section{
Introduction
}
This project calculates quantum $\mathrm{Ca}^{2+}$ interactions with EEG\@.
In this paper,
EEG is synonymous with large-scale neocortical firings during attentional
tasks as measured by large-amplitude electroencephalographic (EEG) recordings.
In this paper,
only very specific calcium ions, $\mathrm{Ca}^{2+}$, are considered,
those arising from regenerative calcium waves generated at tripartite
neuron-astrocyte-neuron synapses.
Indeed, it is important to note that $\mathrm{Ca}^{2+}$ ions, and
specifically
$\mathrm{Ca}^{2+}$ waves, influence many processes in the brain,
but this study focuses on free waves generated at tripartite synapses
because of their calculated direct interactions with large synchronous
neuronal firings.

Section 2 reviews the background of the main model used, Statistical
Mechanics of Neocortical Interactions (SMNI).

Section 3 reviews the code Adaptive Simulated Annealing (ASA), used for optimization
of many systems --- fitting models to real data, e.g.,
fits to EEG data reported here.

Section 4 reviews the development of path-integral codes, PATHINT
and qPATHINT, used for propagation of conditional probabilities and
quantum-mechanical wave-functions, as reported here.

Section 5 gives new results of inclusion of quantum-mechanical interactions
of $\mathrm{Ca}^{2+}$ wave-packets with EEG\@.

Section 6 reviews some applications of this project.

Section 7 gives the conclusion.

The theory and codes for ASA and [q]PATHINT have been well tested across many
disciplines by multiple users.
This particular project most certainly is speculative, but it is testable.
As reported here, fitting such models to EEG tests some aspects of this project.
This is a somewhat indirect path, but not novel to many physics paradigms that are
tested by experiment or computation.
A detailed future path is described in the [q]PATHINT review Section.

While SMNI has been developed since 1981, and been confirmed by many tests,
this evolving model including ionic scales has been part of multiple papers
relatively recently, since 2012.
Classical physics calculations support these extended SMNI models and are consistent
with experimental data.
Quantum physics calculations also support these extended SMNI models and, while they too are
consistent with experimental data, it is quite speculative that they can
persist in neocortex.
Admittedly, it is surprising that detailed calculations continue to support this
model, and so it is worth continued examination it until it is theoretically or
experimentally proven to be false.

\section{
Statistical Mechanics of Neocortical Interactions (SMNI)
}
SMNI has been developed since 1981, scaling aggregate synaptic interactions
to neuronal firings, up to minicolumnar-macrocolumnar columns of
neurons to mesocolumnar dynamics, up to columns of neuronal firings,
up to regional macroscopic sites
\citep{Ingber1981,Ingber1982,Ingber1983,Ingber1984,Ingber1985a,Ingber1994}.

SMNI has calculated
agreement/fits with experimental data from
various aspects of neocortical interactions, e.g.,
properties of short-term memory (STM)
\citep{Ingber2012a},
including its
capacity (auditory $7\pm 2$ and visual $4\pm 2$)
\citep{Ericsson+Chase1982,Zhang+Simon1985},
duration, stability, primacy versus recency rule, as well other
phenomenon, e.g., Hick's law
\citep{Hick1952,Ingber1999,Jensen1987},
interactions within macrocolumns calculating
mental rotation of images, etc
\citep{Ingber1982,Ingber1983,Ingber1984,Ingber1985a,Ingber1994}.
SMNI scaled mesocolumns across neocortical regions
to fit EEG data
\citep{Ingber1997a,Ingber1997b,Ingber2012a}.
Fig. 1 depicts this model
\citep{Ingber1983}.

\pagebreak[4]
\begin{center}
\includegraphics[width=4in]{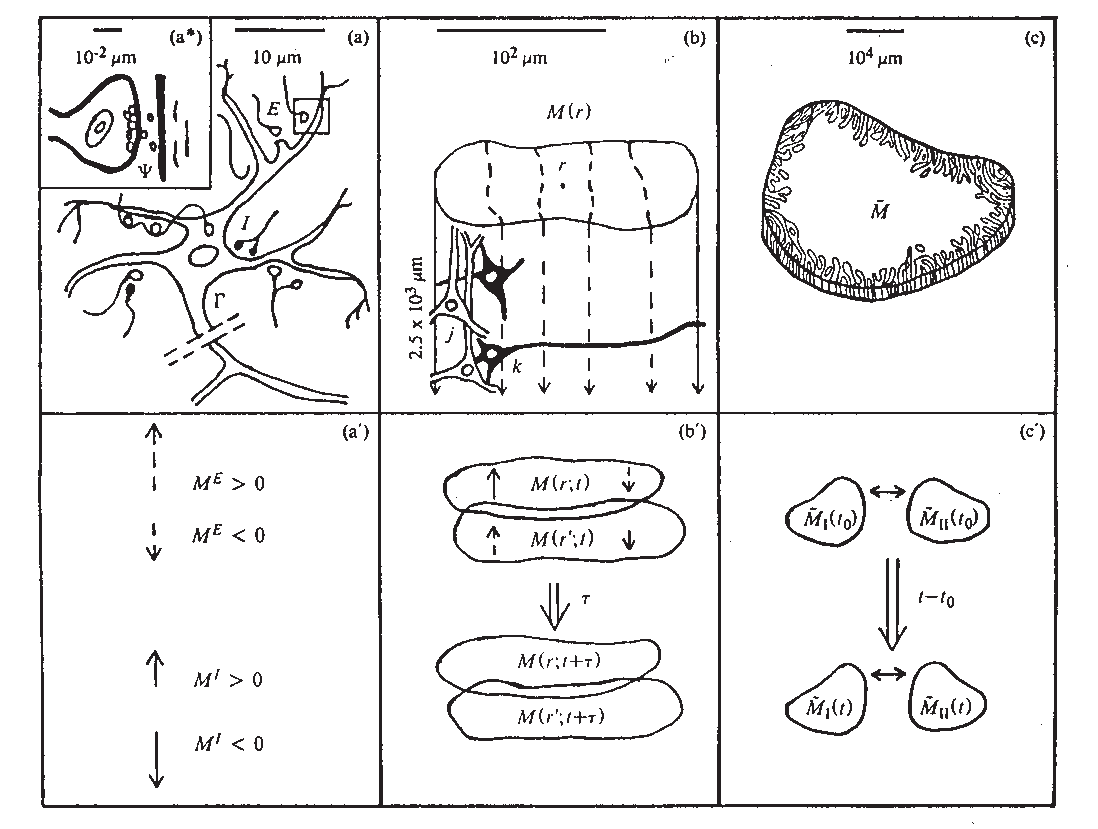}
\end{center}

\begin{quote}

Figure 1 illustrates three SMNI biophysical scales
\citep{Ingber1982,Ingber1983}:
(a)-(a$^{*}$)-(a') microscopic neurons; (b)-(b') mesocolumnar
domains; (c)-(c')
macroscopic regions.
\hfil\\
(a$^{*}$): synaptic inter-neuronal interactions, scaled up to
mesocolumns, phenomenologically described by the mean and
variance of a distribution $\Psi $
(a): intraneuronal transmissions phenomenologically described by
the mean and variance of $\Gamma $
(a'): collective mesocolumnar-averaged inhibitory ($I$)
and excitatory ($E$) neuronal firings $M$
\hfil\\
(b): vertical organization
of minicolumns including their horizontal
layers, yielding a physiological entity, the mesocolumn
(b'): overlapping mesocolumns at locations $r$ and $r'$
from times $t$ and $t+\tau $, $\tau $ on the order of 10 msec
\hfil\\
(c): macroscopic regions of neocortex
arising from many mesocolumnar domains
(c'): regions coupled by long-ranged interactions
\end{quote}

\subsection{
Synaptic Interactions
}
The short-time conditional probability distribution of firing of a given neuron
firing given just-previous firings of other neurons
is calculated from
chemical and electrical intra-neuronal interactions
\citep{Ingber1982,Ingber1983}.
Given its previous interactions with $k$ neurons
within $\tau _{j}$ of 5-10 msec,
the conditional probability that
neuron $j$ fires $(\sigma _{j}=+1)$ or does not fire
$(\sigma _{j}=-1)$ is

\begin{equation*}
p_{\sigma _{j}}=\Gamma \Psi =\frac{\exp (-\sigma _{j}F_{j})}{\exp (F_{j})+\exp (-F_{j})}
\end{equation*}

\begin{equation*}
F_{j}=\frac{V_{j}-\,\sum \limits_{k}\,a_{jk}^{*}v_{jk}}{\big(\pi \,\sum \limits_{k^\prime{}}\,a_{jk^\prime{}}^{*}(v_{jk^\prime{}}^{2}+\phi _{jk^\prime{}}^{2})\big)^{1/2}}
\end{equation*}

\begin{equation}
a_{jk}=\frac{1}{2}A_{|jk|}(\sigma _{k}+1)+B_{jk}
\end{equation}
The contribution
to polarization achieved at an axon given activity at a synapse,
taking into account averaging over different neurons, geometries, etc.,
is given by $\Gamma $, the ``intra-neuronal'' probability distribution.
$\Psi $ is the ``inter-neuronal'' probability distribution,
of thousands of quanta of neurotransmitters released at one
neuron's presynaptic site effecting a (hyper-)polarization at another
neuron's postsynaptic site, taking into account interactions with
neuromodulators, etc.
This development holds for $\Gamma $ Poisson, and for $\Psi $ Poisson or Gaussian.

$V_{j}$ is the depolarization threshold in the somatic-axonal
region.
$v_{jk}$ is the
induced synaptic polarization of $E$ or $I$ type at the axon, and
$\phi _{jk}$ is its variance.
The efficacy $a_{jk}$
is a sum of $A_{jk}$ from the connectivity
between neurons, activated if the impinging $k$-neuron fires,
and $B_{jk}$ from spontaneous background noise.
The efficacy is related to the impedance across synaptic gaps.

\subsection{
Neuronal Interactions
}
Aggregation up to the mesoscopic scale
from the microscopic synaptic scale uses
mesoscopic probability $P$

\begin{equation*}
P=\,\prod \limits_{G}\,P^{G}[M^{G}(r;t+\tau )|M^{\bar{G}}(r^\prime{};t)]
\end{equation*}

\begin{equation}
=\,\sum \limits_{\sigma _{j}}\delta \left (\,\sum \limits_{jE}\,\sigma _{j}-M^{E}(r;t+\tau )\right )\delta \left (\,\sum \limits_{jI}\,\sigma _{j}-M^{I}(r;t+\tau )\right )\prod \limits_{j}\limits^{N}\,p_{\sigma _{j}}
\end{equation}
$M$ represents a mesoscopic scale of columns of $N$ neurons, with subsets
$E$ and $I$, represented by
$p_{q_{i}}$.
The ``delta''-functions $\delta $-constraint
represents an aggregate of many neurons in a column.
$G$ is used to represent excitatory ($E$) and inhibitory ($I$)
contributions.
$\bar{G}$ designates contributions from both $E$ and $I$.

The path integral is derived in terms of mesoscopic Lagrangian $L$.
The short-time distribution of firings in a minicolumn,
given its just previous interactions with all other neurons
in its macrocolumn, is thereby defined.

\subsection{
Columnar Interactions
}
In the prepoint (Ito) representation
the SMNI Lagrangian $L$ is

\begin{equation*}
L=\sum \limits_{G,G^\prime{}}(2N)^{-1}(\dot{M}^{G}-g^{G})g_{GG^\prime{}}(\dot{M}^{G^\prime{}}-g^{G^\prime{}})/(2N\tau )-V^\prime{}
\end{equation*}

\begin{equation*}
g^{G}=-\tau ^{-1}(M^{G}+N^{G}\tanh F^{G})
\end{equation*}

\begin{equation*}
g^{GG^\prime{}}=(g_{GG^\prime{}})^{-1}=\delta _{G}^{G^\prime{}}\tau ^{-1}N^{G}\mathrm{sech}^{2}F^{G}
\end{equation*}

\begin{equation}
g=\det (g_{GG^\prime{}})
\end{equation}
The threshold factor $F^{G}$ is derived as

\begin{equation*}
F^{G}=\sum \limits_{G^\prime{}}\frac{\nu ^{G}+\nu ^{\ddagger E^\prime{}}}{\left ((\pi /2)[(v_{G^\prime{}}^{G})^{2}+(\phi _{G^\prime{}}^{G})^{2}](\delta ^{G}+\delta ^{\ddagger E^\prime{}})\right )^{1/2}}
\end{equation*}

\begin{equation*}
\nu ^{G}=V^{G}-a_{G^\prime{}}^{G}v_{G^\prime{}}^{G}N^{G^\prime{}}-\frac{1}{2}A_{G^\prime{}}^{G}v_{G^\prime{}}^{G}M^{G^\prime{}},\nu ^{\ddagger E^\prime{}}=-a_{E^\prime{}}^{\ddagger E}v_{E^\prime{}}^{E}N^{\ddagger E^\prime{}}-\frac{1}{2}A_{E^\prime{}}^{\ddagger E}v_{E^\prime{}}^{E}M^{\ddagger E^\prime{}}
\end{equation*}

\begin{equation*}
\delta ^{G}=a_{G^\prime{}}^{G}N^{G^\prime{}}+\frac{1}{2}A_{G^\prime{}}^{G}M^{G^\prime{}},\delta ^{\ddagger E^\prime{}}=a_{E^\prime{}}^{\ddagger E}N^{\ddagger E^\prime{}}+\frac{1}{2}A_{E^\prime{}}^{\ddagger E}M^{\ddagger E^\prime{}}
\end{equation*}

\begin{equation}
a_{G^\prime{}}^{G}=\frac{1}{2}A_{G^\prime{}}^{G}+B_{G^\prime{}}^{G},a_{E^\prime{}}^{\ddagger E}=\frac{1}{2}A_{E^\prime{}}^{\ddagger E}+B_{E^\prime{}}^{\ddagger E}
\end{equation}
where
$A_{G^\prime{}}^{G}$ is the columnar-averaged direct synaptic efficacy,
$B_{G^\prime{}}^{G}$ is the columnar-averaged background-noise
contribution to synaptic efficacy.
The ``$^{\ddagger }$'' parameters arise from regional interactions
across many macrocolumns.

\subsection{
SMNI Parameters From Experiments
}
All values of parameters and their bounds are taken from experimental data,
not arbitrarily fit to specific phenomena.

$N^{G}$ =
\{$N^{E}=160$, $N^{I}=60$\} was set for for visual neocortex,
\{$N^{E}=80$, $N^{I}=30$\} was set for all other neocortical regions,
$M^{G^\prime{}}$ and $N^{G^\prime{}}$ in $F^{G}$ are afferent
macrocolumnar firings scaled to efferent minicolumnar firings
by $N/N^{*}\approx 10^{-3}$.
$N^{*}$ is the number of neurons
in a macrocolumn, about $10^{5}$.
$V^\prime{}$ includes nearest-neighbor mesocolumnar interactions.
$\tau $ is usually considered to be on the order of 5-10~ms.

Other values also are consistent with experimental data, e.g.,
$V^{G}=10$~mV,
$v_{G^\prime{}}^{G}=0.1$~mV,
$\phi _{G^\prime{}}^{G}=0.03^{1/2}$~mV.

Nearest-neighbor interactions among columns give dispersion relations
that were used to calculate speeds of mental visual rotation
\citep{Ingber1982,Ingber1983}.

The wave equation cited by EEG theorists, permitting
fits of SMNI to EEG data
\citep{Ingber1995a},
was derived using the variational principle applied to the SMNI Lagrangian.

This creates an audit trail from synaptic parameters
to the statistically
averaged regional Lagrangian.

\subsection{
Previous Applications
}
\subsubsection{
Verification of basic SMNI Hypothesis
}
The core SMNI hypothesis first developed circa 1980
\citep{Ingber1981,Ingber1982,Ingber1983}
is that highly synchronous patterns of neuronal firings in fact process
high-level information.
Only since 2012 has this hypothesis been verified experimentally
\citep{Asher2012,Salazar+Dotson+Bressler+Gray2012}.

\subsubsection{
SMNI Calculations of Short-Term Memory (STM)
}
SMNI calculations agree with observations
\citep{Ingber1982,Ingber1983,Ingber1984,Ingber1985a,Ingber1994,Ingber1995b,Ingber1997a,Ingber1999,Ingber2011,Ingber2012a,Ingber2012b,Ingber2015,Ingber2016a,Ingber2017a,Ingber2017b,Ingber+Pappalepore+Stesiak2014,Nunez+Srinivasan+Ingber2013}:
This list includes:
\begin{quote}
capacity (auditory $7\pm 2$ and visual $4\pm 2$)
\citep{Ingber1984}
\hfil\\
duration
\citep{Ingber1985a}
\hfil\\
stability
\citep{Ingber1985a}
\hfil\\
primacy versus recency rule
\citep{Ingber1985a,Ingber1985b}
\hfil\\
Hick's law (reaction time and $g$ factor)
\citep{Ingber1999}
\hfil\\
nearest-neighbor minicolumnar interactions $=>$ mental rotation of images
\citep{Ingber1982,Ingber1983}
\hfil\\
derivation of basis for EEG
\citep{Ingber1985c,Ingber1995a}
\end{quote}

\subsubsection{
Three Basic SMNI Models
}
Three basic models were developed by slightly
changing the background firing component of the columnar-averaged efficacies
$B_{G^\prime{}}^{G}$ within experimental ranges, which modify
$F^{G}$ threshold factors to yield
in the conditional probability:
\begin{quote}
(a) case EC, dominant excitation subsequent firings
\hfil\\
(b) case IC, inhibitory subsequent firings
\hfil\\
(c) case BC, balanced between EC and IC
\end{quote}

This is consistent with experimental evidence of shifts in background synaptic
activity under conditions of selective attention
\citep{Briggs+Mangun+Usrey2013,Mountcastle+Andersen+Motter1981},
This enables a Centering Mechanism (CM) on case BC, giving $\mathrm{BC}^\prime{}$,
wherein the numerator of $F^{G}$ only has terms proportional to
$M^{E^\prime{}}$, $M^{I^\prime{}}$ and $M^{\ddagger E^\prime{}}$, i.e.,
zeroing other constant terms
by resetting the background parameters $B_{G^\prime{}}^{G}$, still
within experimental ranges.
This brings in a maximum number of minima into the
physical firing $M^{G}$-space,
due to
the minima of the new numerator in being in a parabolic trough defined by

\begin{equation}
A_{E}^{E}M^{E}-A_{I}^{E}M^{I}=0
\end{equation}
about which nonlinearities develop multiple minima identified
with STM phenomena.

In current projects
a Dynamic CM (DCM) model is used,
resetting $B_{G^\prime{}}^{G}$ every few epochs of $\tau $.
Such changes in background synaptic activity on such time scales
are seen during attentional tasks
\citep{Briggs+Mangun+Usrey2013}.

\subsection{
Comparing EEG Testing Data with Training Data
}
Using EEG data from http://physionet.nlm.nih.gov/pn4/erpbci 
\citep{Citi+Poli+Cinel2010,Goldberger+Amaral+Glass+Hausdorff+Ivanov+Mark+Mietus+Moody+Peng+Stanley2000},
SMNI was fit to highly synchronous waves (P300) during attentional tasks,
for each of 12 subjects, it was possible to find 10 Training runs and
10 Testing runs
\citep{Ingber2016a}.

Spline-Laplacian transformations on
the EEG potential $\Phi $ are proportional to the SMNI $M^{G}$ firing variables
at each electrode site.
The electric potential
$\Phi $ is experimentally measured by EEG, not $\mathbf{A}$, but both are
due to the same currents $\mathbf{I}$.
Therefore, $\mathbf{A}$ is linearly proportional to $\Phi $ with a simple
scaling factor included as a parameter in fits to data.
Additional parameterization of background synaptic parameters,
$B_{G^\prime{}}^{G}$ and
$B_{E^\prime{}}^{\ddagger E}$, modify previous work.

The $\mathbf{A}$ model outperformed the no-$\mathbf{A}$
model, where the no-$\mathbf{A}$ model simply has used $\mathbf{A}$-non-dependent
synaptic parameters.
Cost functions with an $|\mathbf{A}|$ model were much worse than either
the $\mathbf{A}$ model or the no-$\mathbf{A}$ model.
Runs with different signs on the drift and on the absolute value of the drift
also gave much higher cost functions than the $\mathbf{A}$ model.

\subsection{
STM PATHINT Calculations
}
\subsubsection{
PATHINT STM
}
The evolution of a Balanced Centered model (BC)
after 500 foldings
of $\Delta t=0.01$, 5 unit of relaxation time $\tau $, exhibits
the existence of ten well developed peaks.
These peaks are identified with possible trappings of firing patterns.

This describes the ``$7\pm 2$'' rule, as calculated by SMNI PATHINT in Fig. 2
\citep{Ingber+Nunez1995}.

\pagebreak[4]
\begin{center}
\includegraphics[width=4in]{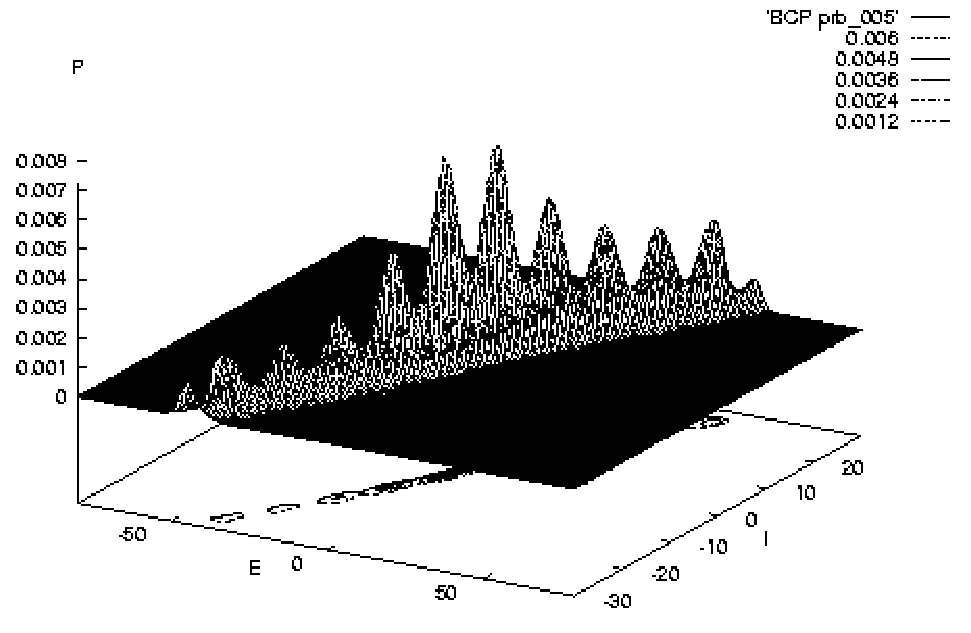}
\end{center}

\begin{quote}
Figure 2 illustrates SMNI STM Model BC at the evolution at 5$\tau $
\citep{Ingber+Nunez1995}.
\end{quote}

\subsubsection{
PATHINT STM Visual
}
The evolution of a Balanced Centered Visual model (BCV)
after 1000 foldings
of $\Delta t=0.01$, 10 unit of relaxation time $\tau $,
exhibits the existence of four well developed peaks.
These peaks are identified with possible trappings of firing patterns.
Other peaks at lower scales are clearly present, numbering on the same
order as in the BC' model, as the strength in the
original peaks dissipates throughout firing space, but these are
much smaller and therefore much less probable to be accessed.

This describes the ``$4\pm 2$'' rule for visual STM,
as calculated by SMNI PATHINT in Fig. 3
\citep{Ingber+Nunez1995}.

\pagebreak[4]
\begin{center}
\includegraphics[width=4in]{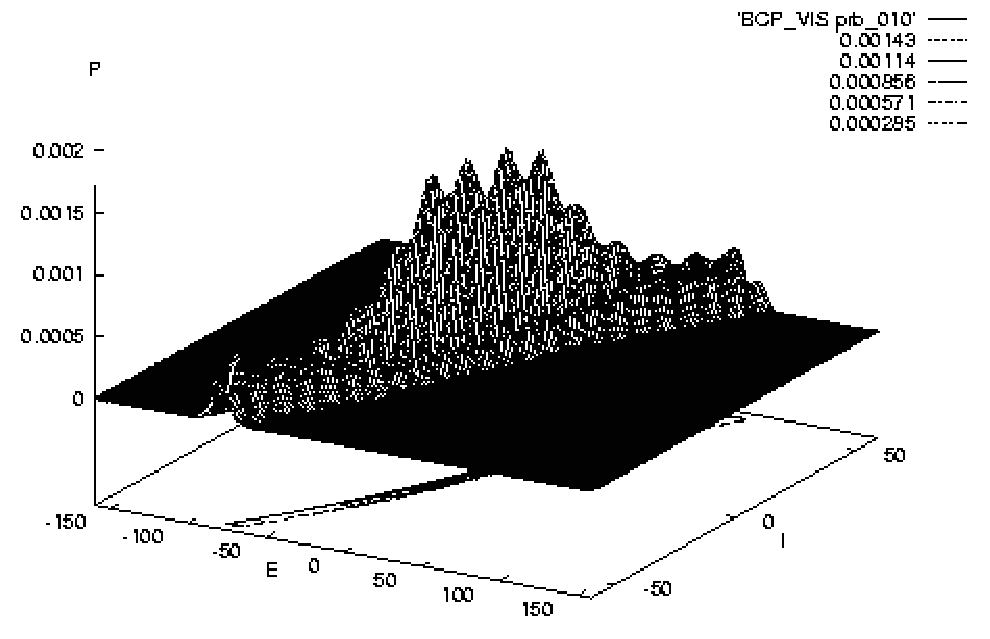}
\end{center}

\begin{quote}
Figure 3 illustrates SMNI STM Model BCV at the evolution at 10$\tau $
\citep{Ingber+Nunez1995}.
\end{quote}

\subsection{
Tripartite Synaptic Interactions
}
The human brain contains over $10^{11}$ cells,
about half of which are neurons.
The other half are glial cells.
Astrocytes comprise a good fraction of glial cells,
possibly the majority.
Many papers examine the influence of astrocytes on synaptic processes
\citep{Agulhon+Petravicz+McMullen+Sweger+Minton+Taves+Casper+Fiacco+McCarthy2008,Araque+Navarrete2010,Banaclocha+Bookkon+Banaclocha2010,Bellinger2005,Innocenti+Parpura+Haydon2000,Pereira+Furlan2009,Reyes+Parpura2009,Scemes+Giaume2006,Volterra+Liaudet+Savtchouk2014}.

http://www.astrocyte.info claims
they are the most numerous cells in the human brain.
Unlike the previous ideology of astrocytes being ``filler'' cells,
they are very active in the central nervous
system and greatly outnumber neurons,

Glutamate release from astrocytes
through a $\mathrm{Ca}^{2+}$-dependent mechanism can activate receptors
at the presynaptic terminals.
Intercellular calcium waves (ICWs)
may travel over hundreds of astrocytes
propagating over many neuronal synapses.
ICWs contribute to control synaptic activity.
Glutamate is released in a regenerative manner, with subsequent cells
that are involved in the calcium wave releasing additional glutamate
\citep{Ross2012}.

$[\mathrm{Ca}^{2+}]$ (concentrations of $\mathrm{Ca}^{2+}$)
affect increased release probabilities at
synaptic sites, by enhancing the release of gliotransmitters.
(Free $\mathrm{Ca}^{2+}$ waves are considered here, not
intracellular astrocyte calcium waves in situ
which also increase neuronal firings.)

These free regenerative $\mathrm{Ca}^{2+}$ waves, arising from
astrocyte-neuron interactions, couple to the magnetic vector potential
$\mathbf{A}$ produced by highly synchronous collective firings, e.g., during
selective attention tasks, as measured by EEG\@.

\subsubsection{
Canonical Momentum $\mathbf{\Pi}=\mathbf{p}+q\mathbf{A}$
}
As derived in the Feynman (midpoint) representation of the path integral,
the canonical momentum, $\mathbf{\Pi}$, defines the
dynamics of a moving particle with
momentum $\mathbf{p}$ in an electromagnetic field.
In SI units,

\begin{equation}
\mathbf{\Pi}=\mathbf{p}+q\mathbf{A}
\end{equation}
where $q=-2e$ for $\mathrm{Ca}^{2+}$,
$e$ is the magnitude of the charge of an electron $=1.6\times 10^{-19}$~C (Coulomb),
and $\mathbf{A}$ is the electromagnetic vector potential.
(In Gaussian units $\mathbf{\Pi}=\mathbf{p}+q\mathbf{A}/c$, where $c$
is the speed of light.)
$\mathbf{A}$ represents three components of a 4-vector.

\subsubsection{
Vector Potential of Wire
}
A columnar firing state is modeled as a wire/neuron with current
$\mathbf{I}$
measured in A $=$ Amperes $=$ C/s,

\begin{equation}
\mathbf{A}(t)=\frac{\mu }{4\pi }\int \frac{dr}{r}\mathbf{I}
\end{equation}
along a length $z$ observed from a
perpendicular distance $r$ from a line of thickness $r_{0}$.
If far-field retardation effects are neglected, this yields

\begin{equation}
\mathbf{A}=\frac{\mu }{4\pi }\mathbf{I}\log \big(\frac{r}{r_{0}}\big)
\end{equation}
where $\mu $ is the
magnetic permeability in vacuum $=4\pi 10^{-7}$ H/m (Henry/meter).
Note the insensitive log dependence on distance.

The contribution to $\mathbf{A}$ includes minicolumnar
lines of current from hundreds to thousands of macrocolumns,
within a region not so large to include many convolutions,
but still contributing to large synchronous bursts of EEG\@.

Electric $\mathbf{E}$ and magnetic $\mathbf{B}$ fields, derivatives of $\mathbf{A}$ with respect to $r$,
do not possess this logarithmic insensitivity
to distance, and therefore they do not
linearly accumulate strength within and across macrocolumns.

Estimates of contributions from
synchronous firings
to P300 measured on the scalp are tens of thousands of macrocolumns
spanning
100 to 100's of cm$^{2}$.
Electric fields generated from a minicolumn may fall
by half within 5-10~mm, the range of several macrocolumns.

There are other possible sources of magnetic vector potentials not described
as wires with currents
\citep{Majhi+Ghosh2018}.
Their net effects plausibly would be included the vector magnetic potential
of net synchrous firings, but not their functional forms as derived here.

\subsubsection{
Effects of Vector Potential on Momenta
}
The momentum $\mathbf{p}$
for a $\mathrm{Ca}^{2+}$ ion with mass $m=6.6\times 10^{-26}$~kg,
speed on the order of 50~$\mu $m/s
to 100~$\mu $m/s,
is on the order of $10^{-30}$~kg-m/s.
Molar concentrations
of $\mathrm{Ca}^{2+}$ waves, comprised of tens of thousands of
free ions representing about 1\% of a released set, most being buffered,
are within a range of about 100~$\mu $m to as much as 250~$\mu $m,
with a duration of more than 500~ms, and with [$\mathrm{Ca}^{2+}$]
ranging from 0.1-5~$\mu $M ($\mu $M = $10^{-3}$~mol/m$^{3}$).

The magnitude of the current is taken from experimental data on dipole
moments $\mathbf{Q}=|\mathbf{I}|\mathbf{\hat{z}}$ where $\mathbf{\hat{z}}$ is the direction
of the current $\mathbf{I}$ with the dipole spread over $z$.
$\mathbf{Q}$ ranges from 1~pA-m = $10^{-12}$~A-m for a
pyramidal neuron
\citep{Murakami+Okada2006},
to $10^{-9}$~A-m for larger neocortical mass
\citep{Nunez+Srinivasan2006}.
These currents give rise to $q\mathbf{A}\approx10^{-28}$~kg-m/s.
The velocity of a $\mathrm{Ca}^{2+}$ wave can be $\approx$20-50~$\mu $m/s.
In neocortex, a typical $\mathrm{Ca}^{2+}$ wave of 1000 ions, with total
mass $m=6.655\times 10^{-23}$~kg
times a speed of $\approx$20-50~$\mu $m/s, gives
$\mathbf{p}\approx10^{-27}$~kg-m/s.

Taking $10^{4}$ synchronous firings in a macrocolumn, leads to a
dipole moment $|\mathbf{Q}|=10^{-8}$~A-m.
Taking $z$ to be $10^{2}\mu $m $=10^{-4}$~m,
a couple of neocortical layers, gives
$|q\mathbf{A}|\approx 2\times 10^{-19}\times 10^{-7}\times 10^{-8}/10^{-4}$
= $10^{-28}$~kg-m/s,

\subsubsection{
Reasonable Estimates
}
Estimates used here for $\mathbf{Q}$ come from experimental data.
These include shielding and material effects.
When coherent activity among many macrocolumns associated with STM
is considered,
$|\mathbf{A}|$ may be much larger.
Since $\mathrm{Ca}^{2+}$ waves influence synaptic activity,
there is direct coherence between these waves and the activity of $\mathbf{A}$.

Classical physics calculates $q\mathbf{A}$ from macroscopic EEG
to be on the order of $10^{-28}$~kg-m/s, while the momentum $\mathbf{p}$
of a $\mathrm{Ca}^{2+}$ ion is on the order of $10^{-30}$~kg-m/s.
This numerical comparison demonstrates the possible importance of the influence
of $\mathbf{A}$ on $\mathbf{p}$ at classical scales.

This project fits the SMNI model to EEG data.
Direct calculations in classical and quantum physics support the concepts
presented here, e.g., that ionic calcium momentum-wave effects among
neuron-astrocyte-neuron tripartite synapses modify background SMNI parameters
and create feedback between ionic/quantum and macroscopic scales
\citep{Ingber2012a,Ingber2012b,Ingber2015,Ingber2016a,Ingber2017a,Ingber+Pappalepore+Stesiak2014,Nunez+Srinivasan+Ingber2013}.

\subsection{
Model of Models (MOM)
}
Deep Learning (DL) has invigorated AI approaches to parsing data in complex systems,
often to develop control processes of these systems.
A couple of decades ago, Neural Net AI approaches fell out of favor when concerns
were apparent that such approaches offered little guidance to explain the ``why'' or
``how'' such algorithms worked to process data, e.g., contexts which were deemed important
to deal with future events and outliers, etc.

The success of DL has overshadowed these concerns.
However, that should not diminish their importance, especially if such systems
are placed in positions to affect lives and human concerns;
humans are ultimately responsible for structures they build.

An approach to dealing with these concerns can be called Model of Models (MOM).
An argument in favor of MOM is that humans over thousands of years have
developed models of reality across many disciplines, e.g., ranging over
Physics, Biology, Mathematics, Economics, etc.

A good use of DL might be to process data for a given system in terms of
a collection of models, then again use DL to process the models over the
same data to determine a superior model of models (MOM).
Eventually, large DL (quantum) machines could possess a database of hundreds or
thousands of models across many disciplines, and directly find the best
(hybrid) MOM for a given system.

In particular, SMNI offers a reasonable model upon which to further develop
MOM, wherein multiple scales of observed interactions are developed.
This is just one example of how physics modeling and computational physics
can be used to better understand complex systems.

\subsubsection{
Ideas by Statistical Mechanics
}
A project sympathetic to this MOM context was proposed as Ideas by Statistical
Mechanics (ISM)
\citep{Ingber2006a,Ingber2007,Ingber2008}.
using ASA
\citep{Ingber1989,Ingber1993a,Ingber2012c}
to fit parameters of a generic nonlinear multivariate colored-noise
Gaussian-Markovian short-time conditional probability distribution
to data, useful for many systems.

Models developed using ASA have been applied in many contexts across many systems
\citep{Ingber1993b},
including applications to neural networks
\citep{Atiya+Parlos+Ingber2003}.

Many of these ASA applications have used Ordinal representations
of features, to permit parameterization of their inclusion into models,
quite similar in spirit to DL.

ASA can be used again in the expanded context of MOM.
This is suggested as a first step in a new discipline to which MOM
is to be applied, to help develop a range of parameters useful for DL,
as DL by itself may get stuck in non-ideal local minima of the
importance-sampled space.
Then, after a reasonable range of models is found, DL can take over
to permit much more efficient and accurate development of MOM for a
given discipline/system.

\section{
Adaptive Simulated Annealing (ASA) Algorithm
}
\subsection{
Importance Sampling
}
Nonlinear and/or stochastic systems often require importance-sampling algorithms
to scan or to fit parameters.
Methods of simulated annealing (SA) are often used.
Proper annealing (not ``quenching'') possesses a proof of finding the
deepest minimum in searches.

The ASA code is open-source software, and can be downloaded and used without any cost or registration
at https://www.ingber.com/\#ASA 
\citep{Ingber1993a,Ingber2012c}.

This algorithm fits empirical data to a cost function over a
$D$-dimensional parameter space,
adapting for varying sensitivities of parameters during the fit.

This ASA algorithm is
faster than fast Cauchy annealing, which has schedule $T_{i}=T_{0}/k$,
and much faster than Boltzmann annealing, which has schedule $T_{i}=T_{0}/\ln k$
\citep{Ingber1989}.

\subsection{
Outline of ASA Algorithm
}
For parameters

\begin{equation*}
\alpha _{k}^{i}\in[A_{i},B_{i}]
\end{equation*}

sampling with the random variable $x^{i}$

\begin{equation*}
x^{i}\in[-1,1]
\end{equation*}

\begin{equation*}
\alpha _{k+1}^{i}=\alpha _{k}^{i}+x^{i}(B_{i}-A_{i})
\end{equation*}

the default generating function is

\begin{equation*}
g_{T}(x)=\prod \limits_{i=1}\limits^{D}\,\frac{1}{2\,\ln (1+1/T_{i})(|x^{i}|+T_{i})}\equiv \prod \limits_{i=1}\limits^{D}\,g_{T}^{i}(x^{i})
\end{equation*}

in terms of parameter ``temperatures''

\begin{equation}
T_{i}=T_{i0}\exp (-c_{i}k^{1/D})
\end{equation}
The default ASA uses the same type of annealing schedule for
the acceptance function $h$ as used for the generating function $g$.

All default functions in ASA can be overridden with user-defined functions.

\subsection{
ASA Applications
}
The ASA code
\citep{Ingber1993a}
and the original Very Fast Simulated Reannealing (VFSR) code
\citep{Ingber1989}
have been used by many researchers, including the author in a range of disciplines:
\begin{quote}
chaotic systems
\citep{Ingber+Srinivasan+Nunez1996}
\hfil\\
combat simulations
\citep{Ingber1993c,Ingber1998a}
\hfil\\
financial systems: bonds, equities, futures, options
\citep{Ingber1990,Ingber1996a,Ingber2000,Ingber2005,Ingber+Chen+Mondescu+Muzzall+Renedo2001,Ingber+Mondescu2003}
\hfil\\
neuroscience
\citep{Ingber1991,Ingber1992,Ingber1996b,Ingber1997a,Ingber1998b,Ingber2006b,Ingber2009a,Ingber2009b,Ingber2012a,Ingber2012b,Ingber2013,Ingber2015,Ingber+Nunez1995,Ingber+Nunez2010,Ingber+Pappalepore+Stesiak2014,Ingber+Srinivasan+Nunez1996,Nunez+Srinivasan+Ingber2013}
\hfil\\
optimization \it{}per se\rm{}
\citep{Atiya+Parlos+Ingber2003,Ingber1989,Ingber1993b,Ingber1996c,Ingber2012c,Ingber+Rosen1992}
\end{quote}

\section{
Path-Integral Algorithms PATHINT and qPATHINT
}
\subsection{
Path Integral in Stratonovich (Midpoint) Representation
}
The path integral in the Feynman (midpoint) representation is used to
examine discretization issues in time-dependent nonlinear systems
\citep{Langouche+Roekaerts+Tirapegui1979,Langouche+Roekaerts+Tirapegui1982,Schulman1981}.
(N.b. $g^{\dagger }$ in $DM$ implies a prepoint evaluation.)
Unless explicitly stated, the Einstein summation convention is used which
implies repeated indices signify summation; bars $|\ldots |$ imply no summation.

\begin{equation*}
P[M_{t}|M_{t_{0}}]dM(t)=\int \ldots \int \,DM\exp \left (-\min \,\int \limits_{t_{0}}\limits^{t}\,dt^\prime{}L\right )\delta \big(M(t_{0})=M_{0}\big)\delta \big(M(t)=M_{t}\big)
\end{equation*}

\begin{equation*}
DM=\lim \limits_{u\rightarrow \infty }\,\prod \limits_{\rho =1}\limits^{u+1}\,g^{\dagger 1/2}\,\prod \limits_{G}\,(2\pi \theta )^{-1/2}dM_{\rho }^{G}
\end{equation*}

\begin{equation*}
L(\dot{M}^{G},M^{G},t)=\frac{1}{2}(\dot{M}^{G}-h^{G})g_{GG^\prime{}}(\dot{M}^{G^\prime{}}-h^{G^\prime{}})+\frac{1}{2}h^{G}_{;G}+R/6-V
\end{equation*}

\begin{equation*}
\dot{M}^{G}(t)\rightarrow M_{\rho +1}^{G}-M_{\rho }^{G},M^{G}(t)\rightarrow \frac{1}{2}(M_{\rho +1}^{G}+M_{\rho }^{G}),[\ldots ]_{,G}=\frac{\partial [\ldots ]}{\partial M^{G}}
\end{equation*}

\begin{equation*}
h^{G}=g^{G}-\frac{1}{2}g^{-1/2}(g^{1/2}g^{GG^\prime{}})_{,G^\prime{}},h^{G}_{;G}=h_{,G}^{G}+\Gamma _{GF}^{F}h^{G}=g^{-1/2}(g^{1/2}h^{G})_{,G}
\end{equation*}

\begin{equation*}
g_{GG^\prime{}}=(g^{GG^\prime{}})^{-1},g=\det (g_{GG^\prime{}})
\end{equation*}

\begin{equation*}
\Gamma _{JK}^{F}\equiv g^{LF}[JK,L]=g^{LF}(g_{JL,K}+g_{KL,J}-g_{JK,L})
\end{equation*}

\begin{equation*}
R=g^{JL}R_{JL}=g^{JL}g^{JK}R_{FJKL}
\end{equation*}

\begin{equation}
R_{FJKL}=\frac{1}{2}(g_{FK,JL}-g_{JK,FL}-g_{FL,JK}+g_{JL,FK})+g_{MN}(\Gamma _{FK}^{M}\Gamma _{JL}^{N}-\Gamma _{FL}^{M}\Gamma _{JK}^{N})
\end{equation}
Non-constant diffusions add terms to
drifts, and a Riemannian-curvature potential $R/6$ is induced for dimension
$>1$
in the Stratonovich/Feynman discretization.

\subsection{
Path Integral in Ito (Prepoint) Representation
}
In the Ito (prepoint) representation:

\begin{equation*}
P[M_{t}|M_{t_{0}}]dM(t)=\int \ldots \int DM\exp \big(-\min \,\int \limits_{t_{0}}\limits^{t}\,dt^\prime{}L\big)\delta \big(M(t_{0})=M_{0}\big)\delta \big(M(t)=M_{t}\big)
\end{equation*}

\begin{equation*}
DM=\lim \limits_{u\rightarrow \infty }\,\prod \limits_{\rho =1}\limits^{u+1}\,g^{1/2}\,\prod \limits_{G}\,(2\pi \Delta t)^{-1/2}dM_{\rho }^{G}
\end{equation*}

\begin{equation*}
L(\dot{M}^{G},M^{G},t)=\frac{1}{2}(\dot{M}^{G}-g^{G})g_{GG^\prime{}}(\dot{M}^{G^\prime{}}-g^{G^\prime{}})+R/6
\end{equation*}

\begin{equation*}
\dot{M}^{G}(t)\rightarrow M_{\rho +1}^{G}-M_{\rho }^{G},M^{G}(t)\rightarrow M_{\rho }^{G}
\end{equation*}

\begin{equation}
(g_{GG^\prime{}})=(g^{GG^\prime{}})^{-1},g=\det (g_{GG^\prime{}})
\end{equation}
Here the diagonal diffusions are $g^{|GG|}$
and the drifts are $g^{G}$.

\subsection{
Path-Integral Riemannian Geometry
}
The midpoint derivation derives a Riemannian geometry
with metric defined by the inverse of the covariance matrix

\begin{equation}
g_{GG^\prime{}}=(g^{GG^\prime{}})^{-1}
\end{equation}
and where $R$ is the Riemannian curvature

\begin{equation}
R=g^{JL}R_{JL}=g^{JL}g^{JK}R_{FJKL}
\end{equation}
An Ito prepoint discretization for the same probability distribution $P$
gives a simpler algebraic form,

\begin{equation*}
M(\bar{t}_{s})=M(t_{s})
\end{equation*}

\begin{equation}
{{L}}=\frac{1}{2}(dM^{G}/dt-g^{G})g_{GG^\prime{}}(dM^{G^\prime{}}/dt-g^{G^\prime{}})-V
\end{equation}
but the Lagrangian ${{L}}$ so specified does not satisfy a
variational principle useful for moderate to large noise.
Its variational principle is only useful in the weak-noise limit.
This often means that finer meshes are required.

\subsection{
Three Approaches Are Mathematically Equivalent
}
Three basic different approaches are mathematically equivalent:
\begin{quote}
(a) Fokker-Planck/Chapman-Kolmogorov partial-differential
equations
\hfil\\
(b) Langevin coupled stochastic-differential equations
\hfil\\
(c) Lagrangian or Hamiltonian path-integrals
\end{quote}
All three are described here as many researchers are familiar with at least
one of these approaches to complex systems.

The path-integral approach is useful to define
intuitive physical variables from the Lagrangian $L$
in terms of underlying variables $M^{G}$:

\begin{equation*}
\mathrm{Momentum}:\:\Pi ^{G}=\frac{\partial L}{\partial (\partial M^{G}/\partial t)}
\end{equation*}

\begin{equation*}
\mathrm{Mass}:\:g_{GG^\prime{}}=\frac{\partial L}{\partial (\partial M^{G}/\partial t)\partial (\partial M^{G^\prime{}}/\partial t)}
\end{equation*}

\begin{equation*}
\mathrm{Force}:\:\frac{\partial L}{\partial M^{G}}
\end{equation*}

\begin{equation}
F=ma:\:\delta L=0=\frac{\partial L}{\partial M^{G}}-\frac{\partial }{\partial t}\frac{\partial L}{\partial (\partial M^{G}/\partial t)}
\end{equation}
Differentiation especially of noisy systems often introduces more noise.
The path-integral often gives superior numerical performance
because integration is a smoothing process.

\subsubsection{
Stochastic Differential Equation (SDE)
}
The Stratonovich (midpoint discretized) Langevin equations can be
analyzed in terms of
the Wiener process $dW^{i}$.
This can be developed with
Gaussian noise $\eta ^{i}=dW^{i}/dt$, with some care
taken in the limit of small $dt$.

\begin{equation*}
dM^{G}=f^{G}\big(t,M(t)\big)dt+\hat{g}_{i}^{G}\big(t,M(t)\big)dW^{i}
\end{equation*}

\begin{equation*}
\dot{M}^{G}(t)=f^{G}\big(t,M(t)\big)+\hat{g}_{i}^{G}\big(t,M(t)\big)\eta ^{i}(t)
\end{equation*}

\begin{equation*}
dW^{i}\rightarrow \eta ^{i}dt
\end{equation*}

\begin{equation*}
M= \{ M^{G};G=1,\ldots ,\Lambda  \} 
\end{equation*}

\begin{equation*}
\eta = \{ \eta ^{i};i=1,\ldots ,N \} 
\end{equation*}

\begin{equation*}
\dot{M}^{G}=dM^{G}/dt
\end{equation*}

\begin{equation}
<\eta ^{j}(t)>_{\eta }=0,<\eta ^{j}(t),\eta ^{j^\prime{}}(t^\prime{})>_{\eta }=\delta ^{jj^\prime{}}\delta (t-t^\prime{})
\end{equation}
$\eta ^{i}$ represents Gaussian white noise.

\subsubsection{
Partial Differential Equation (PDE)
}
The Fokker-Planck, sometimes defines as Chapman-Kolmogorov,
partial differential equation is:

\begin{equation*}
P_{,t}=\frac{1}{2}(g^{GG^\prime{}}P)_{,GG^\prime{}}-(g^{G}P)_{,G}+VP
\end{equation*}

\begin{equation*}
P=<P_{\eta }>_{\eta }
\end{equation*}

\begin{equation*}
g^{G}=f^{G}+\frac{1}{2}\hat{g}_{i}^{G^\prime{}}\hat{g}_{i,G^\prime{}}^{G}
\end{equation*}

\begin{equation*}
g^{GG^\prime{}}=\hat{g}_{i}^{G}\hat{g}_{i}^{G^\prime{}}
\end{equation*}

\begin{equation}
(\ldots )_{,G}=\partial (\ldots )/\partial M^{G}
\end{equation}
$g^{G}$ replaces $f^{G}$ in the SDE if the
Ito (prepoint discretized) calculus is used.
If boundary conditions are added as Lagrange multipliers, these
enter as a ``potential'' $V$
creating a Schrodinger-type equation.

\subsection{
PATHINT Applications
}
Path integrals and PATHINT have been applied across several disciplines, including
combat simulations
\citep{Ingber+Fujio+Wehner1991},
neuroscience
\citep{Ingber1994,Ingber2017b,Ingber+Nunez1995,Ingber+Nunez2010},
finance
\citep{Ingber2000,Ingber2016b,Ingber2017a,Ingber2017b,Ingber2017c,Ingber+Chen+Mondescu+Muzzall+Renedo2001,Ingber+Wilson2000},
and other nonlinear systems
\citep{Ingber1995c,Ingber1998a,Ingber+Srinivasan+Nunez1996}.

\subsection{
PATHINT/qPATHINT Code
}
qPATHINT is an N-dimensional code developed to calculate the propagation of quantum
variables in the presence of shocks.
Many real systems propagate
in the presence of sudden changes of state dependent on time.
qPATHINT is based on the classical-physics code, PATHINT, which has been
useful in several systems across several disciplines.
Applications have been made to
SMNI and
Statistical Mechanics of Financal Markets (SMFM)
\citep{Ingber2017a,Ingber2017b,Ingber2017c}.

To numerically calculate the path integral for serial changes in time,
standard Monte Carlo techniques generally are not useful.
PATHINT was originally developed for this purpose.
The PATHINT C code of about 7500 lines of code
using the GCC C-compiler was rewritten to use double complex variables instead of
double variables, and
further developed for arbitrary N dimensions, creating qPATHINT.
The outline of the code is described here for classical or quantum systems,
using generic coordinates $q$
\citep{Ingber2016b,Ingber2017a,Ingber2017c}.

The distribution (probabilities
for classical systems, wave-functions for quantum systems)
can be numerically approximated to a high degree of accuracy
using a histogram procedure, developing
sums of rectangles
of height $P_{i}$ and width $\Delta q^{i}$
at points $q^{i}$.

\subsubsection{
Shocks
}
Many real-world systems propagate in the presence of continual ``shocks''.

In SMNI, collisions occur via regenerative $\mathrm{Ca}^{2+}$ waves.
There also are interactions with changing $\mathbf{A}$ due to
changing highly synchronous neuronal firings.

In SMFM applications, shocks occur due to future dividends, changes in interest rates,
changes in asset distributions, etc.

\subsubsection{
PATHINT/qPATHINT Histograms
}
A one-dimensional path-integral in variable $q$ in the prepoint Ito discretization
is developed in terms of the kernel/propagator $G$,
for each of its intermediate integrals, as

\begin{equation*}
P(q;t+\Delta t)=\,\int dq^\prime{}[g^{1/2}(2\pi \Delta t)^{-1/2}\exp (-L\,\Delta t)]P(q^\prime{};t)=\,\int dq^\prime{}G(q,q^\prime{};\Delta t)P(q^\prime{};t)
\end{equation*}

\begin{equation*}
P(q;t)=\sum \limits_{i=1}\limits^{N}\pi (q-q^{i})P_{i}(t)
\end{equation*}

\begin{equation}
\pi (q-q^{i})=1\,,\:(q^{i}-\frac{1}{2}\Delta q^{i-1})\le q\le (q^{i}+\frac{1}{2}\Delta q^{i});0\,,\:\mathrm{otherwise}
\end{equation}
This yields

\begin{equation*}
P_{i}(t+\Delta t)=T_{ij}(\Delta t)P_{j}(t)
\end{equation*}

\begin{equation}
T_{ij}(\Delta t)=\frac{2}{\Delta q^{i-1}+\Delta q^{i}}\,\int \limits_{q^{i}-\Delta q^{i-1}/2}\limits^{q^{i}+\Delta q^{i}/2}dq\,\int \limits_{q^{j}-\Delta q^{j-1}/2}\limits^{q^{j}+\Delta q^{j}/2}dq^\prime{}G(q,q^\prime{};\Delta t)
\end{equation}
$T_{ij}$ is a banded matrix representing the Gaussian nature
of the short-time probability centered about the drift.

Several projects have used this algorithm
\citep{Ingber+Nunez1995,Ingber+Srinivasan+Nunez1996,Ingber+Wilson1999,Wehner+Wolfer1983a,Wehner+Wolfer1983b,Wehner+Wolfer1987}.
Special 2-dimensional codes were developed for specific projects
in Statistical Mechanics of Combat (SMC), SMNI and SMFM
\citep{Ingber2000,Ingber+Fujio+Wehner1991,Ingber+Nunez1995}.

\subsubsection{
Meshes For [q]PATHINT
}
Explicit dependence of $L$ on time $t$ can be included.
The mesh $\Delta q^{i}$
is strongly dependent on diagonal elements of the
diffusion matrix, e.g.,

\begin{equation}
\Delta q^{i}\approx (\Delta tg^{|ii|})^{1/2}
\end{equation}
This constrains the dependence of the covariance of each
variable to be a (nonlinear) function of that variable
to present a rectangular underlying mesh.
Since integration is inherently a smoothing process
\citep{Ingber1990},
coarser meshes are used relative to the corresponding stochastic
differential equation(s)
\citep{Wehner+Wolfer1983a}.

By considering the contributions to the first and second moments,
conditions on the time and variable meshes can be derived.
Thus $\Delta t$ can be measured by the diffusion divided by the
square of the drift.

\subsection{
Lessons Learned From SMFM and SMNI
}
SMNI qPATHINT has emphasized the requirement of broad-banded kernels for oscillatory quantum states.

SMFM PATHTREE, and its derived qPATHTREE, is a different options code, based on path-integral error analyses,
permitting a new very fast binary calculation, also applied to nonlinear
time-dependent systems
\citep{Ingber+Chen+Mondescu+Muzzall+Renedo2001}.
However, in contrast to the present PATHINT/qPATHINT code that has been
generalized to N dimensions, currently an SMFM [q]PATHTREE is only a binary tree with $\mathrm{J}=1$ and cannot be effectively applied to quantum oscillatory systems
\citep{Ingber2016b,Ingber2017a,Ingber2017c}.

\subsubsection{
Calculations At Each Node At Each Time Slice
}
SMFM [q]PATHINT for (American) financial options: Calculate at each node of each time slice --- back in time.

SMNI [q]PATHINT: Calculate at each node of each time slice --- forward in time.

At each node of each time slice,
a proposed algorithm is to
calculate quantum-scale $\mathrm{Ca}^{2+}$ wave-packet (2-way) interactions with
macroscopic large-scale EEG/$\mathbf{A}$.
This entails algorithms:
\begin{quote}
PATHINT using the Classical SMNI Lagrangian
\hfil\\
qPATHINT using the Quantum $\mathrm{Ca}^{2+}$ wave-packet Lagrangian
\end{quote}
\hfil\\
Sync in time during P300 attentional tasks.
\hfil\\
Time/phase relations between classical and quantum systems may be important.
\hfil\\
ASA-fit synchronized classical-quantum PATHINT-qPATHINT model to EEG data.
\hfil\\
$\mathbf{A}$ is determined experimentally from EEG, and includes all synaptic
background $B_{G^\prime{}}^{G}$ effects.

\section{
Results Including Quantum Scales
}
The wave function $\psi _{\mathrm{e}}$ describing the interaction of
$\mathbf{A}$ with $\mathbf{p}$ of $\mathrm{Ca}^{2+}$ wave packets
was derived in closed form
from the Feynman representation of the path integral using path-integral
techniques
\citep{Schulten1999},
modified here to include $\mathbf{A}$.

\begin{equation*}
\psi _{\mathrm{e}}(t)=\int d\mathbf{r}_{0}\psi _{0}\psi _{F}=\left [\frac{1-i \hbar t/(m\Delta \mathbf{r}^{2})}{1+i \hbar t/(m\Delta \mathbf{r}^{2})}\right ]^{1/4}\left [\pi \Delta \mathbf{r}^{2}\mathrm{ \{ }1+[ \hbar t/(m\Delta \mathbf{r}^{2})]^{2}\mathrm{ \} }\right ]^{-1/4}
\end{equation*}

\begin{equation*}
\times \exp \left [-\frac{[\mathbf{r}-(\mathbf{p}_{0}+q\mathbf{A})t/m]^{2}}{2\Delta \mathbf{r}^{2}}\frac{1-i \hbar t/(m\Delta \mathbf{r}^{2})}{1+[ \hbar t/(m\Delta \mathbf{r}^{2})]^{2}}+i\frac{\mathbf{p}_{0} \cdot \mathbf{r}}{\hbar }-i\frac{(\mathbf{p}_{0}+q\mathbf{A})^{2}t}{2 \hbar m}\right ]
\end{equation*}

\begin{equation*}
\psi _{F}(t)=\int \frac{d\mathbf{p}}{2\pi  \hbar }\exp \left [\frac{i}{\hbar }\left (\mathbf{p}(\mathbf{r}-\mathbf{r}_{0})-\frac{\mathbf{\Pi}^{2}t}{(2m)}\right )\right ]=\left [\frac{m}{2\pi i \hbar t}\right ]^{1/2}\exp \left [\frac{im(\mathbf{r}-\mathbf{r}_{0}-q\mathbf{A}t/m)^{2}}{2 \hbar t}-\frac{i(q\mathbf{A})^{2}t}{2m \hbar }\right ]
\end{equation*}

\begin{equation}
\psi _{0}=\psi (\mathbf{r}_{0},t=0)=\left (\frac{1}{\pi \Delta \mathbf{r}^{2}}\right )^{1/4}\exp \left (-\frac{\mathbf{r}_{0}^{2}}{2\Delta \mathbf{r}^{2}}+i\frac{\mathbf{p}_{0} \cdot \mathbf{r}_{0}}{\hbar }\right )
\end{equation}
where $\psi _{0}$ is the initial Gaussian packet, $\psi _{F}$ is the free-wave
evolution operator,
$ \hbar $ is the Planck constant,
$q$ is the electronic charge of $\mathrm{Ca}^{2+}$ ions,
$m$ is the mass of a wave-packet of 1000 $\mathrm{Ca}^{2+}$ ions,
$\Delta \mathbf{r}^{2}$ is the spatial variance of the wave-packet,
the initial momentum is $\mathbf{p}_{0}$,
and the evolving canonical momentum is $\mathbf{\Pi}=\mathbf{p}+q\mathbf{A}$.
Detailed calculations show that $\mathbf{p}$ of the
$\mathrm{Ca}^{2+}$ wave packet and $q\mathbf{A}$ of the EEG field
make about equal contributions to $\mathbf{\Pi}$
\citep{Ingber2015}.

\subsection{
SMNI + $\mathrm{Ca}^{2+}$ wave-packet
}
Tripartite influence on synaptic $B_{G^\prime{}}^{G}$ is measured by the ratio of packet's $<\mathbf{p}(t)>_{\psi *\psi }$
to $<\mathbf{p}_{0}(t_{0})>_{\psi *\psi }$ at the onset of each attentional task.
Here $<>_{\psi *\psi }$ is taken over $\psi _{\mathrm{e}}^{*}\,\psi _{\mathrm{e}}$.

\begin{equation}
<\mathbf{p}>_{\psi *\psi }=m\frac{<\mathbf{r}>_{\psi *\psi }}{t-t_{0}}=\frac{q\mathbf{A}+\mathbf{p}_{0}}{m^{1/2}|\Delta \mathbf{r}|}\left (\frac{( \hbar t)^{2}+(m\Delta \mathbf{r}^{2})^{2}}{\hbar t+m\Delta \mathbf{r}^{2}}\right )^{1/2}
\end{equation}
$\mathbf{A}$ changes slower than $\mathbf{p}$, so static approximation of $\mathbf{A}$
used to derive $\psi _{\mathrm{e}}$ and $<\mathbf{p}>_{\psi *\psi }$ is reasonable to
use within P300 EEG epochs, resetting $t=0$ at the onset of each
classical EEG measurement (1.953 ms apart), using the current $\mathbf{A}$.
This permits tests of interactions across scales in a classical context.

\subsection{
Supercomputer Resources
}
The XSEDE.org University of California San Diego (UCSD) supercomputer resource
is Comet, described at https://portal.xsede.org/sdsc-comet .

About 1000 hours of supercomputer CPUs
are required for an ASA fit of SMNI to the same EEG data used previously, i.e.,
from
http://physionet.nlm.nih.gov/pn4/erpbci 
\citep{Citi+Poli+Cinel2010,Goldberger+Amaral+Glass+Hausdorff+Ivanov+Mark+Mietus+Moody+Peng+Stanley2000},
using mostly the same codes used previously
\citep{Ingber2016a}.
Many such sets of runs are required.
Including quantum processes will take even longer.

\subsection{
Results Using $<\mathbf{p}>_{\psi *\psi }$
}
$<\mathbf{p}>_{\psi *\psi }$ was used in classical-physics SMNI fits to EEG data using ASA.
Runs using 1M or 100K generated states gave results not much different.
Training with ASA used 100K generated states over 12 subjects with and without $\mathbf{A}$, followed by 1000 generated states with the simplex local code contained with ASA.
Training and Testing runs on XSEDE.org for this project has taken an equivalent of several months of CPU on the XSEDE.org UCSD platform Comet.
These calculations
use one additional parameter across
all EEG regions to weight the contribution to synaptic background $B_{G^\prime{}}^{G}$.
$\mathbf{A}$ is taken to be proportional to the currents measured by EEG,
i.e., firings $M^{G}$.
Otherwise, the ``zero-fit-parameter'' SMNI philosophy was enforced,
wherein parameters are picked from
experimentally determined values or
within experimentally determined ranges
\citep{Ingber1984}.

As with previous studies using this data, results sometimes give
Testing cost functions less than the Training cost functions.
This reflects on great differences in data, likely from great differences
in subjects' contexts, e.g., possibly
due to subjects' STM strategies only
sometimes including effects calculated here.
Further tests of these multiple-scale models with
more EEG data are required, and with the PATHINT-qPATHINT coupled algorithm described previously.

Table 1 gives recent results on such tests.
Cost functions are the effective Action, $A_{eff}$,
which is $L\,\Delta t-\log (\mathrm{prefactor})$, where the prefactor multiplier of the exponential arises from the normalization
of the short-time conditional probability distribution and $L\,\Delta t$ is the
argument of the exponential factor.
Eq. (3) defines the Lagrangian $L$, and the normalization is defined in $DM$ in Eq. (11).

\begin{table}\small
\captionsetup{width=0.8\textwidth}
\centering
\caption{Column 1 is the subject number; the other columns are cost functions.
Columns 2 and 3 are {no-\textbf{A}} model's Training (TR0) and Testing (TE0).
Columns 4 and 5 are \textbf{A} model's Training (TR\textbf{A}) and Testing (TE\textbf{A}).
Columns 6 and 7 are switched {no-\textbf{A}} model's Training (sTR0) and Testing (sTE0).
Columns 8 and 9 are switched \textbf{A} model's Training (sTR\textbf{A}) and Testing (sTE\textbf{A}).}
\label{tab:t1}
\begin{tabular}{|l|D{.}{.}{-1}D{.}{.}{-1}|D{.}{.}{-1}D{.}{.}{-1}|D{.}{.}{-1}D{.}{.}{-1}|D{.}{.}{-1}D{.}{.}{-1}|}
\hline
\multicolumn{1}{|c}{Sub} &
\multicolumn{1}{|c}{TR0} &
\multicolumn{1}{c}{TE0} &
\multicolumn{1}{|c}{TR\textbf{A}} &
\multicolumn{1}{c}{TE\textbf{A}} &
\multicolumn{1}{|c}{sTR0} &
\multicolumn{1}{c}{sTE0} &
\multicolumn{1}{|c}{sTR\textbf{A}} &
\multicolumn{1}{c|}{sTE\textbf{A}} \\
\hline
s01 & 85.75 & 121.23 & 84.76 & 121.47 & 120.48 & 86.59 & 119.23 & 87.06 \\
s02 & 70.80 & 51.21 & 68.63 & 56.51 & 51.10 & 70.79 & 49.36 & 74.53 \\
s03 & 61.37 & 79.81 & 59.83 & 78.79 & 79.20 & 61.50 & 75.22 & 79.17 \\
s04 & 52.25 & 64.20 & 50.09 & 66.99 & 63.55 & 52.83 & 63.27 & 64.60 \\
s05 & 67.28 & 72.04 & 66.53 & 72.78 & 71.38 & 67.83 & 69.60 & 68.13 \\
s06 & 84.57 & 69.72 & 80.22 & 64.13 & 69.09 & 84.67 & 61.74 & 114.21 \\
s07 & 68.66 & 78.65 & 68.28 & 86.13 & 78.48 & 68.73 & 75.57 & 69.58 \\
s08 & 46.58 & 43.81 & 44.24 & 49.38 & 43.28 & 47.27 & 42.89 & 63.09 \\
s09 & 47.22 & 24.88 & 46.90 & 25.77 & 24.68 & 47.49 & 24.32 & 49.94 \\
s10 & 53.18 & 33.33 & 53.33 & 36.97 & 33.14 & 53.85 & 30.32 & 55.78 \\
s11 & 43.98 & 51.10 & 43.29 & 52.76 & 50.95 & 44.47 & 50.25 & 45.85 \\
s12 & 45.78 & 45.14 & 44.38 & 46.08 & 44.92 & 46.00 & 44.45 & 46.56 \\
\hline
\end{tabular}
\end{table}

\subsection{
Quantum Zeno Effects
}
The quantum-mechanical wave function of the wave packet
was shown to ``survive''
overlaps after multiple collisions, due to their regenerative processes
during the observed long durations of hundreds of ms.
Thus, $\mathrm{Ca}^{2+}$ waves
may support a Zeno or ``bang-bang'' effect
which may promote long coherence times
\citep{Burgarth+Facchi+Nakazato+Pascazio+Yuasa2018,Facchi+Lidar+Pascazio2004,Facchi+Pascazio2008,Giacosa+Pagliara2014,Kozlowski+Caballero-Benitez+Mekhov2015,Muller+Gherardini+Caruso2016,Patil+Chakram+Vengalattore2015,Wu+Wang+Yi2012,Zhang+Ai+Li+Xu+Sun2014}.

Of course, the Zeno/``bang-bang'' effect may exist only in special contexts,
since decoherence among particles is known to be very fast, e.g., faster
than phase-damping of macroscopic classical particles colliding
with quantum particles
\citep{Preskill2015}.

The wave may be perpetuated
by the constant collisions of ions as they enter
and leave the wave packet due to the regenerative
collisions
by the Zeno/``bang-bang'' effect.
qPATHINT
can calculate the coherence stability of the wave
due to serial shocks.

\subsubsection{
Survival of Wave Packet
}
In momentum space, the wave packet
$\phi (\mathbf{p},t)$ is considered as being
``kicked'' from $\mathbf{p}$ to $\mathbf{p}+\delta \mathbf{p}$.
Assume that random repeated kicks of $\delta \mathbf{p}$ result in
$<\delta \mathbf{p}>\approx 0$, and that each kick keeps the variance
$\Delta (\mathbf{p}+\delta \mathbf{p})^{2}\approx \Delta (\mathbf{p})^{2}$.
Then, the overlap integral at the moment $t$ of a typical kick
between the new and old state is

\begin{equation*}
<\phi ^{*}(\mathbf{p}+\delta \mathbf{p},t)|\phi (\mathbf{p},t)>=\exp \left (\frac{i\kappa +\rho }{\sigma }\right )
\end{equation*}

\begin{equation*}
\kappa =8\delta \mathbf{p}\Delta \mathbf{p}^{2} \hbar m(q\mathbf{A}+\mathbf{p}_{0})t-4(\delta \mathbf{p}\Delta \mathbf{p}^{2}t)^{2}
\end{equation*}

\begin{equation*}
\rho =-(\delta \mathbf{p} \hbar m)^{2}
\end{equation*}

\begin{equation}
\sigma =8(\Delta \mathbf{p} \hbar m)^{2}
\end{equation}
where $\phi (\mathbf{p}+\delta \mathbf{p},t)$ is the normalized wave function in
$\mathbf{p}+\delta \mathbf{p}$ momentum space.
A crude estimate is obtained of the survival time amplitude $A(t)$ and
survival probability $p(t)$
\citep{Facchi+Pascazio2008},

\begin{equation*}
A(t)=<\phi ^{*}(\mathbf{p}+\delta \mathbf{p},t)|\phi (\mathbf{p},t)>
\end{equation*}

\begin{equation}
p(t)=|A(t)|^{2}
\end{equation}
These numbers yield:

\begin{equation}
<\phi ^{*}(\mathbf{p}+\delta \mathbf{p},t)|\phi (\mathbf{p},t)>=\exp \big(i(1.67\times 10^{-1}t-1.15\times 10^{-2}t^{2})-1.25\times 10^{-7}\big)
\end{equation}
Even many small repeated kicks do not appreciably affect the real part of $\phi $,
and these projections do not appreciably destroy the original wave packet,
giving a survival probability per kick as $p(t)\approx \exp (-2.5\times 10^{-7})\approx 1-2.5\times 10^{-7}$.

The time-dependent phase terms are sensitive to
times of tenths of a sec.
These times are prominent in
STM and in synchronous neural firings.
Therefore, $\mathbf{A}$ effects on $\mathrm{Ca}^{2+}$ wave functions may maximize
their influence on STM
at frequencies consistent with synchronous EEG during STM\@.

All these calculations support this model, in contrast to other models
of quantum brain processes without such specific calculations and support
\citep{Hagan+Hameroff+Tuszynski2002,Hameroff+Penrose2013,McKemmish+Reimers+McKenzie+Mark+Hush2009}.

\section{
Quantum Applications
}
\subsection{
Nano-Robotic Applications
}
There is the possibility of carrying pharmaceutical products
in nanosystems that could affect unbuffered $\mathrm{Ca}^{2+}$ waves in neocortex
\citep{Ingber2015}.
A $\mathrm{Ca}^{2+}$-wave momentum-sensor could act like a piezoelectric device.

At the onset of a $\mathrm{Ca}^{2+}$ wave (on the order of 100's of ms),
a change of momentum can be on the order of $10^{-30}$~kg-m/s
for a typical $\mathrm{Ca}^{2+}$ ion.
A $\mathrm{Ca}^{2+}$ wave packet of 1000 ions with onset time of 1~ms,
exerts a force on the order of $10^{-24}$~N
(1~N $\equiv $ 1~Newton = 1~kg-m/s$^{2}$).
A nano-robot would be attracted to this site,
depositing chemicals/drugs that interact with the regenerative $\mathrm{Ca}^{2+}$-wave
process.

An area of the receptor of the nanosystem of
1~nm$^{2}$
would require
pressure sensitivity of
$10^{-6}$~Pa (1~Pa = 1~pascal = 1~N/m$^{2}$).

The nano-roboot could be switched on/off at a regional/columnar level
by sensitivity to local electric/magnetic fields.
Highly synchronous firings during STM processes
can be affected by these
piezoelectric nanosystems which affect background/noise
efficacies
via control of $\mathrm{Ca}^{2+}$ waves.
In turn, this would affect the influence of of $\mathrm{Ca}^{2+}$ waves
via the vector potential $\mathbf{A}$, etc.

\subsection{
Free Will
}
There is
interest in researching possible quantum influences on highly
synchronous neuronal firings relevant to STM to understand
connections to consciousness and ``Free Will'' (FW)
\citep{Ingber2016a,Ingber2016b}.

If experimental evidence is gained of quantum-level
processes of tripartite synaptic interactions
with large-scale synchronous neuronal firings, then FW may be
established using the Conway-Kochen quantum no-clone ``Free Will Theorem'' (FWT)
\citep{Conway+Kochen2006,Conway+Kochen2009}.

The essence of FWT is that,
since quantum states cannot be cloned,
a $\mathrm{Ca}^{2+}$ quantum wave-packet may not
generate a state proven to have previously existed.
As explained by the authors
\citep{Conway+Kochen2006,Conway+Kochen2009},
experimenters have specific choices in selecting measurements,
which are shared by (twinned) particles, including the choice of any random number
generator that might be used to aid such choices.
The authors maintain that their proof and description of quantum measurements used
is general enough to rule out classical randomness, and that classical
determinism cannot be supported by such processes as exist in the quantum world.

\section{
Conclusion
}
The SMNI
model has demonstrated it is faithful to experimental data, for
EEG recordings under
STM experimental paradigms.
qPATHINT permits an inclusion of quantum scales in the
multiple-scale SMNI model, by evolving $\mathrm{Ca}^{2+}$
wave-packets with momentum $\mathbf{p}$,
including serial shocks,
interacting with the magnetic vector potential
$\mathbf{A}$ derived from EEG data,
marching forward in time lock-step with experimental
EEG data.
This presents a time-dependent propagation of interacting
quantum and classical scales.

This quantum path-integral algorithm
with serial random shocks will be further studied
as it can be used for many quantum systems.

\section*{Acknowledgment}
The author thanks the
Extreme Science and Engineering Discovery Environment (XSEDE.org),
for supercomputer grants since February 2013, starting with
``Electroencephalographic field influence on calcium momentum waves'',
one under PHY130022 and two under TG-MCB140110.
The current grant under TG-MCB140110,
``Quantum path-integral qPATHTREE and qPATHINT algorithms'', was started
in 2017, and renewed through December 2018.
XSEDE grants have spanned several projects described in
https://www.ingber.com/lir\_computational\_physics\_group.html .

\clearpage
\section*{}
\bibliographystyle{IEEEtranSN}\bibliography{lingber}

\begin{thebibliography}{102}
\providecommand{\natexlab}[1]{#1}
\providecommand{\url}[1]{#1}
\csname url@samestyle\endcsname
\providecommand{\newblock}{\relax}
\providecommand{\bibinfo}[2]{#2}
\providecommand{\BIBentrySTDinterwordspacing}{\spaceskip=0pt\relax}
\providecommand{\BIBentryALTinterwordstretchfactor}{4}
\providecommand{\BIBentryALTinterwordspacing}{\spaceskip=\fontdimen2\font plus
\BIBentryALTinterwordstretchfactor\fontdimen3\font minus
  \fontdimen4\font\relax}
\providecommand{\BIBforeignlanguage}[2]{{%
\expandafter\ifx\csname l@#1\endcsname\relax
\typeout{** WARNING: IEEEtranSN.bst: No hyphenation pattern has been}%
\typeout{** loaded for the language `#1'. Using the pattern for}%
\typeout{** the default language instead.}%
\else
\language=\csname l@#1\endcsname
\fi
#2}}
\providecommand{\BIBdecl}{\relax}
\BIBdecl

\bibitem[A.~Pereira and Furlan(2009)]{Pereira+Furlan2009}
J.~A.~Pereira and F.~Furlan, ``On the role of synchrony for neuron-astrocyte
  interactions and perceptual conscious processing,'' \emph{Journal of
  Biological Physics}, vol.~35, no.~4, pp. 465--480, 2009.

\bibitem[Agulhon et~al.(2008)Agulhon, Petravicz, McMullen, Sweger, Minton,
  Taves, Casper, Fiacco, and
  McCarthy]{Agulhon+Petravicz+McMullen+Sweger+Minton+Taves+Casper+Fiacco+McCarthy2008}
C.~Agulhon, J.~Petravicz, A.~McMullen, E.~Sweger, S.~Minton, S.~Taves,
  K.~Casper, T.~Fiacco, and K.~McCarthy, ``What is the role of astrocyte
  calcium in neurophysiology?'' \emph{Neuron}, vol.~59, pp. 932--946, 2008.

\bibitem[Araque and Navarrete(2010)]{Araque+Navarrete2010}
A.~Araque and M.~Navarrete, ``Glial cells in neuronal network function,''
  \emph{Philosophical Transactions of The Royal Society B}, pp. 2375--2381,
  2010.

\bibitem[Asher(2012)]{Asher2012}
J.~Asher, ``{Brain's} code for visual working memory deciphered in monkeys
  {NIH}-funded study,'' NIH, Bethesda, MD, Tech. Rep. NIH Press Release, 2012,
  {}\url{http://www.nimh.nih.gov/news/science-news/2012/in-sync-brain-waves-hold-memory-of-objects-just-seen.shtml}.

\bibitem[Atiya et~al.(2003)Atiya, Parlos, and Ingber]{Atiya+Parlos+Ingber2003}
A.~Atiya, A.~Parlos, and L.~Ingber, ``A reinforcement learning method based on
  adaptive simulated annealing,'' in \emph{Proceedings International Midwest
  Symposium on Circuits and Systems (MWCAS), December 2003}.\hskip 1em plus
  0.5em minus 0.4em\relax Cairo, Egypt: IEEE CAS, 2003,
  {}\url{https://www.ingber.com/asa03_reinforce.pdf}.

\bibitem[Banaclocha et~al.(2010)Banaclocha, Bookkon, and
  Banaclocha]{Banaclocha+Bookkon+Banaclocha2010}
M.~Banaclocha, I.~Bookkon, and H.~Banaclocha, ``Long-term memory in brain
  magnetite,'' \emph{Medical Hypotheses}, vol.~74, no.~2, pp. 254--257, 2010.

\bibitem[Bellinger(2005)]{Bellinger2005}
S.~Bellinger, ``Modeling calcium wave oscillations in astrocytes,''
  \emph{Neurocomputing}, vol.~65, no.~66, pp. 843--850, 2005.

\bibitem[Briggs et~al.(2013)Briggs, Mangun, and Usrey]{Briggs+Mangun+Usrey2013}
F.~Briggs, G.~Mangun, and W.~Usrey, ``Attention enhances synaptic efficacy and
  the signal-to-noise ratio in neural circuits,'' \emph{Nature}, vol. 000, pp.
  1--5, 2013, {}\url{https://doi.org/10.1038/nature12276}.

\bibitem[Burgarth et~al.(2018)Burgarth, Facchi, Nakazato, Pascazio, and
  Yuasa]{Burgarth+Facchi+Nakazato+Pascazio+Yuasa2018}
D.~Burgarth, P.~Facchi, H.~Nakazato, S.~Pascazio, and K.~Yuasa, ``Quantum zeno
  dynamics from general quantum operations,'' Aberystwyth U., Aberystwyth, UK,
  Tech. Rep. arXiv:1809.09570 [quant-ph], 2018,
  {}\url{https://arxiv.org/pdf/1510.04857.pdf}.

\bibitem[Citi et~al.(2010)Citi, Poli, and Cinel]{Citi+Poli+Cinel2010}
L.~Citi, R.~Poli, and C.~Cinel, ``Documenting, modelling and exploiting {P300}
  amplitude changes due to variable target delays in {Donchin's} speller,''
  \emph{Journal of Neural Engineering}, vol.~7, no. 056006, pp. 1--21, 2010,
  {}\url{https://doi.org/10.1088/1741-2560/7/5/056006}.

\bibitem[Conway and Kochen(2006)]{Conway+Kochen2006}
J.~Conway and S.~Kochen, ``The free will theorem,'' Princeton U, Princeton, NJ,
  Tech. Rep. arXiv:quant-ph/0604079 [quant-ph], 2006,
  {}\url{https://arxiv.org/pdf/quant-ph/0604079.pdf}.

\bibitem[Conway and Kochen(2009)]{Conway+Kochen2009}
J.~Conway and S.~Kochen, ``The strong free will theorem,'' \emph{Notices of the American
  Mathematical Society}, vol.~56, no.~2, pp. 226--232, 2009.

\bibitem[Ericsson and Chase(1982)]{Ericsson+Chase1982}
K.~Ericsson and W.~Chase, ``Exceptional memory,'' \emph{American Scientist},
  vol.~70, pp. 607--615, 1982.

\bibitem[Facchi and Pascazio(2008)]{Facchi+Pascazio2008}
P.~Facchi and S.~Pascazio, ``Quantum zeno dynamics: mathematical and physical
  aspects,'' \emph{Journal of Physics A}, vol.~41, no. 493001, pp. 1--45, 2008.

\bibitem[Facchi et~al.(2004)Facchi, Lidar, and
  Pascazio]{Facchi+Lidar+Pascazio2004}
P.~Facchi, D.~Lidar, and S.~Pascazio, ``Unification of dynamical decoupling and
  the quantum zeno effect,'' \emph{Physical Review A}, vol.~69, no. 032314, pp.
  1--6, 2004.

\bibitem[Giacosa and Pagliara(2014)]{Giacosa+Pagliara2014}
G.~Giacosa and G.~Pagliara, ``Quantum zeno effect by general measurements,''
  \emph{Physical Review A}, vol. 052107, pp. 1--5, 2014.

\bibitem[Goldberger et~al.(2000)Goldberger, Amaral, Glass, Hausdorff, Ivanov,
  Mark, Mietus, Moody, Peng, and
  Stanley]{Goldberger+Amaral+Glass+Hausdorff+Ivanov+Mark+Mietus+Moody+Peng+Stanley2000}
A.~Goldberger, L.~Amaral, L.~Glass, J.~Hausdorff, P.~Ivanov, R.~Mark,
  J.~Mietus, G.~Moody, C.-K. Peng, and H.~Stanley, ``{PhysioBank,}
  {PhysioToolkit,} and {PhysioNet:} components of a new research resource for
  complex physiologic signals,'' \emph{Circulation}, vol. 101, no.~23, pp.
  e215--e220, 2000,
  {}\url{http://circ.ahajournals.org/cgi/content/full/101/23/e215}.

\bibitem[Hagan et~al.(2002)Hagan, Hameroff, and
  Tuszynski]{Hagan+Hameroff+Tuszynski2002}
S.~Hagan, S.~Hameroff, and J.~Tuszynski, ``Quantum computation in brain
  microtubules: Decoherence and biological feasibility,'' \emph{Physical Review
  E}, vol.~65, no. 061901, pp. 1--11, 2002,
  {}\url{https://doi.org/10.1103/PhysRevE.65.061901}.

\bibitem[Hameroff and Penrose(2013)]{Hameroff+Penrose2013}
S.~Hameroff and R.~Penrose, ``Consciousness in the universe: A review of the
  {'Orch} {OR'} theory,'' \emph{Physics of Life Reviews}, vol. 403, pp. 1--40,
  2013, {}\url{https://doi.org/10.1016/j.plrev.2013.08.002}.

\bibitem[Hick(1952)]{Hick1952}
W.~Hick, ``On the rate of gains of information,'' \emph{Quarterly Journal
  Experimental Psychology}, vol.~34, no.~4, pp. 1--33, 1952.

\bibitem[Ingber(1981)]{Ingber1981}
L.~Ingber, ``Towards a unified brain theory,'' \emph{Journal Social Biological
  Structures}, vol.~4, pp. 211--224, 1981,
  {}\url{https://www.ingber.com/smni81_unified.pdf}.

\bibitem[Ingber(1982)]{Ingber1982}
L.~Ingber, ``Statistical mechanics of neocortical interactions. i. basic
  formulation,'' \emph{Physica D}, vol.~5, pp. 83--107, 1982,
  {}\url{https://www.ingber.com/smni82_basic.pdf}.

\bibitem[Ingber(1983)]{Ingber1983}
L.~Ingber, ``Statistical mechanics of neocortical interactions. dynamics of
  synaptic modification,'' \emph{Physical Review A}, vol.~28, pp. 395--416,
  1983, {}\url{https://www.ingber.com/smni83_dynamics.pdf}.

\bibitem[Ingber(1984)]{Ingber1984}
L.~Ingber, ``Statistical mechanics of neocortical interactions. derivation of
  short-term-memory capacity,'' \emph{Physical Review A}, vol.~29, pp.
  3346--3358, 1984, {}\url{https://www.ingber.com/smni84_stm.pdf}.

\bibitem[Ingber(1985a)]{Ingber1985a}
L.~Ingber, ``Statistical mechanics of neocortical interactions: Stability and
  duration of the {7+}-2 rule of short-term-memory capacity,'' \emph{Physical
  Review A}, vol.~31, pp. 1183--1186, 1985a,
  {}\url{https://www.ingber.com/smni85_stm.pdf}.

\bibitem[Ingber(1985b)]{Ingber1985b}
L.~Ingber, ``Towards clinical applications of statistical mechanics of neocortical
  interactions,'' \emph{Innovations Technology Biology Medicine}, vol.~6, pp.
  753--758, 1985b.

\bibitem[Ingber(1985c)]{Ingber1985c}
L.~Ingber, ``Statistical mechanics of neocortical interactions. {EEG} dispersion
  relations,'' \emph{IEEE Transactions in Biomedical Engineering}, vol.~32, pp.
  91--94, 1985c, {}\url{https://www.ingber.com/smni85_eeg.pdf}.

\bibitem[Ingber(1989)]{Ingber1989}
L.~Ingber, ``Very fast simulated re-annealing,'' \emph{Mathematical Computer
  Modelling}, vol.~12, no.~8, pp. 967--973, 1989,
  {}\url{https://www.ingber.com/asa89_vfsr.pdf}.

\bibitem[Ingber(1990)]{Ingber1990}
L.~Ingber, ``Statistical mechanical aids to calculating term structure models,''
  \emph{Physical Review A}, vol.~42, no.~12, pp. 7057--7064, 1990,
  {}\url{https://www.ingber.com/markets90_interest.pdf}.

\bibitem[Ingber(1991)]{Ingber1991}
L.~Ingber, ``Statistical mechanics of neocortical interactions: A scaling paradigm
  applied to electroencephalography,'' \emph{Physical Review A}, vol.~44,
  no.~6, pp. 4017--4060, 1991, {}\url{https://www.ingber.com/smni91_eeg.pdf}.

\bibitem[Ingber(1992)]{Ingber1992}
L.~Ingber, ``Generic mesoscopic neural networks based on statistical mechanics of
  neocortical interactions,'' \emph{Physical Review A}, vol.~45, no.~4, pp.
  R2183--R2186, 1992, {}\url{https://www.ingber.com/smni92_mnn.pdf}.

\bibitem[Ingber(1993a)]{Ingber1993a}
L.~Ingber, ``Adaptive simulated annealing {(ASA)},'' Caltech Alumni Association,
  Pasadena, CA, Tech. Rep. Global optimization C-code, 1993a,
  {}\url{https://www.ingber.com/\#ASA-CODE}.

\bibitem[Ingber(1993b)]{Ingber1993b}
L.~Ingber, ``Simulated annealing: Practice versus theory,'' \emph{Mathematical
  Computer Modelling}, vol.~18, no.~11, pp. 29--57, 1993b,
  {}\url{https://www.ingber.com/asa93_sapvt.pdf}.

\bibitem[Ingber(1993c)]{Ingber1993c}
L.~Ingber, ``Statistical mechanics of combat and extensions,'' in \emph{Toward a
  Science of Command, Control, and Communications}, C.~Jones, Ed.\hskip 1em
  plus 0.5em minus 0.4em\relax Washington, D.C.: American Institute of
  Aeronautics and Astronautics, 1993c, pp. 117--149, {}ISBN 1-56347-068-3.
  \url{https://www.ingber.com/combat93_c3sci.pdf}.

\bibitem[Ingber(1994)]{Ingber1994}
L.~Ingber, ``Statistical mechanics of neocortical interactions: Path-integral
  evolution of short-term memory,'' \emph{Physical Review E}, vol.~49, no.~5B,
  pp. 4652--4664, 1994, {}\url{https://www.ingber.com/smni94_stm.pdf}.

\bibitem[Ingber(1995a)]{Ingber1995a}
L.~Ingber, ``Statistical mechanics of multiple scales of neocortical
  interactions,'' in \emph{Neocortical Dynamics and Human EEG Rhythms},
  P.~Nunez, Ed.\hskip 1em plus 0.5em minus 0.4em\relax New York, NY: Oxford
  University Press, 1995a, pp. 628--681, {}ISBN 0-19-505728-7.
  \url{https://www.ingber.com/smni95_scales.pdf}.

\bibitem[Ingber(1995b)]{Ingber1995b}
L.~Ingber, ``Statistical mechanics of neocortical interactions: Constraints on 40
  hz models of short-term memory,'' \emph{Physical Review E}, vol.~52, no.~4,
  pp. 4561--4563, 1995b, {}\url{https://www.ingber.com/smni95_stm40hz.pdf}.

\bibitem[Ingber(1995c)]{Ingber1995c}
L.~Ingber, ``Path-integral evolution of multivariate systems with moderate
  noise,'' \emph{Physical Review E}, vol.~51, no.~2, pp. 1616--1619, 1995c,
  {}\url{https://www.ingber.com/path95_nonl.pdf}.

\bibitem[Ingber(1996a)]{Ingber1996a}
L.~Ingber, ``Canonical momenta indicators of financial markets and neocortical
  {EEG},'' in \emph{Progress in Neural Information Processing}, S.-I. Amari,
  L.~Xu, I.~King, and K.-S. Leung, Eds.\hskip 1em plus 0.5em minus 0.4em\relax
  New York: Springer, 1996a, pp. 777--784, {}Invited paper to the 1996
  International Conference on Neural Information Processing (ICONIP'96), Hong
  Kong, 24-27 September 1996. ISBN 981-3083-05-0.
  \url{https://www.ingber.com/markets96_momenta.pdf}.

\bibitem[Ingber(1996b)]{Ingber1996b}
L.~Ingber, ``Statistical mechanics of neocortical interactions: Multiple scales of
  {EEG},'' in \emph{Frontier Science in EEG: Continuous Waveform Analysis
  (Electroencephal. clin. Neurophysiol. Suppl. 45)}, R.~Dasheiff and
  D.~Vincent, Eds.\hskip 1em plus 0.5em minus 0.4em\relax Amsterdam: Elsevier,
  1996b, pp. 79--112, {}Invited talk to Frontier Science in EEG Symposium, New
  Orleans, 9 Oct 1993. ISBN 0-444-82429-4.
  \url{https://www.ingber.com/smni96_eeg.pdf}.

\bibitem[Ingber(1996c)]{Ingber1996c}
L.~Ingber, ``Adaptive simulated annealing {(ASA):} lessons learned,''
  \emph{Control and Cybernetics}, vol.~25, no.~1, pp. 33--54, 1996c, {}Invited
  paper to Control and Cybernetics on Simulated Annealing Applied to
  Combinatorial Optimization. \url{https://www.ingber.com/asa96_lessons.pdf}.

\bibitem[Ingber(1997a)]{Ingber1997a}
L.~Ingber, ``Statistical mechanics of neocortical interactions: Applications of
  canonical momenta indicators to electroencephalography,'' \emph{Physical
  Review E}, vol.~55, no.~4, pp. 4578--4593, 1997a,
  {}\url{https://www.ingber.com/smni97_cmi.pdf}.

\bibitem[Ingber(1997b)]{Ingber1997b}
L.~Ingber, \emph{{EEG} Database}.\hskip 1em plus 0.5em minus 0.4em\relax Irvine,
  CA: UCI Machine Learning Repository, 1997b,
  {}\url{http://archive.ics.uci.edu/ml/datasets/EEG+Database}.

\bibitem[Ingber(1998a)]{Ingber1998a}
L.~Ingber, ``Data mining and knowledge discovery via statistical mechanics in
  nonlinear stochastic systems,'' \emph{Mathematical Computer Modelling},
  vol.~27, no.~3, pp. 9--31, 1998a,
  {}\url{https://www.ingber.com/path98_datamining.pdf}.

\bibitem[Ingber(1998b)]{Ingber1998b}
L.~Ingber, ``Statistical mechanics of neocortical interactions: Training and
  testing canonical momenta indicators of {EEG},'' \emph{Mathematical Computer
  Modelling}, vol.~27, no.~3, pp. 33--64, 1998b,
  {}\url{https://www.ingber.com/smni98_cmi_test.pdf}.

\bibitem[Ingber(1999)]{Ingber1999}
L.~Ingber, ``Statistical mechanics of neocortical interactions: Reaction time
  correlates of the g factor,'' \emph{Psycholoquy}, vol.~10, no. 068, 1999,
  {}Invited commentary on The g Factor: The Science of Mental Ability by Arthur
  Jensen. \url{http://www.cogsci.ecs.soton.ac.uk/cgi/psyc/newpsy?10.068}.

\bibitem[Ingber(2000)]{Ingber2000}
L.~Ingber, ``High-resolution path-integral development of financial options,''
  \emph{Physica A}, vol. 283, no. 3-4, pp. 529--558, 2000,
  {}\url{https://www.ingber.com/markets00_highres.pdf}.

\bibitem[Ingber(2005)]{Ingber2005}
L.~Ingber, ``Trading in risk dimensions {(TRD)},'' Lester Ingber Research,
  Ashland, OR, Tech. Rep. Report 2005:TRD, 2005,
  {}\url{https://www.ingber.com/markets05_trd.pdf}.

\bibitem[Ingber(2006a)]{Ingber2006a}
L.~Ingber, ``Ideas by statistical mechanics {(ISM)},'' Lester Ingber Research,
  Ashland, OR, Tech. Rep. Report 2006:ISM, 2006a,
  {}\url{https://www.ingber.com/smni06_ism.pdf}.

\bibitem[Ingber(2006b)]{Ingber2006b}
L.~Ingber, ``Statistical mechanics of neocortical interactions: Portfolio of
  physiological indicators,'' Lester Ingber Research, Ashland, OR, Tech. Rep.
  Report 2006:PPI, 2006b, {}\url{https://www.ingber.com/smni06_ppi.pdf}.

\bibitem[Ingber(2007)]{Ingber2007}
L.~Ingber, ``Ideas by statistical mechanics {(ISM)},'' \emph{Journal Integrated
  Systems Design and Process Science}, vol.~11, no.~3, pp. 31--54, 2007,
  {}Special Issue: Biologically Inspired Computing.

\bibitem[Ingber(2008)]{Ingber2008}
L.~Ingber, ``{AI} and ideas by statistical mechanics {(ISM)},'' in
  \emph{Encyclopedia of Artificial Intelligence}, J.~Rabunal, J.~Dorado, and
  A.~Pazos, Eds.\hskip 1em plus 0.5em minus 0.4em\relax New York: Information
  Science Reference, 2008, pp. 58--64, {}ISBN 978-1-59904-849-9.

\bibitem[Ingber(2009a)]{Ingber2009a}
L.~Ingber, ``Statistical mechanics of neocortical interactions: Portfolio of
  physiological indicators,'' \emph{The Open Cybernetics Systemics Journal},
  vol.~3, no.~14, pp. 13--26, 2009a,
  {}\url{https://doi.org/10.2174/1874110x00903010013}.

\bibitem[Ingber(2009b)]{Ingber2009b}
L.~Ingber, ``Statistical mechanics of neocortical interactions: Nonlinear columnar
  electroencephalography,'' \emph{NeuroQuantology Journal}, vol.~7, no.~4, pp.
  500--529, 2009b, {}\url{https://www.ingber.com/smni09_nonlin_column_eeg.pdf}.

\bibitem[Ingber(2011)]{Ingber2011}
L.~Ingber, ``Computational algorithms derived from multiple scales of neocortical
  processing,'' in \emph{Pointing at Boundaries: Integrating Computation and
  Cognition on Biological Grounds}, J.~A.~Pereira, E.~Massad, and N.~Bobbitt,
  Eds.\hskip 1em plus 0.5em minus 0.4em\relax New York: Springer, 2011, pp.
  1--13, {}Invited Paper. \url{https://doi.org/10.1007/s12559-011-9105-4}.

\bibitem[Ingber(2012a)]{Ingber2012a}
L.~Ingber, ``Columnar {EEG} magnetic influences on molecular development of
  short-term memory,'' in \emph{Short-Term Memory: New Research}, G.~Kalivas
  and S.~Petralia, Eds.\hskip 1em plus 0.5em minus 0.4em\relax Hauppauge, NY:
  Nova, 2012a, pp. 37--72, {}Invited Paper.
  \url{https://www.ingber.com/smni11_stm_scales.pdf}.

\bibitem[Ingber(2012b)]{Ingber2012b}
L.~Ingber, ``Influence of macrocolumnar {EEG} on ca waves,'' \emph{Current
  Progress Journal}, vol.~1, no.~1, pp. 4--8, 2012b,
  {}\url{https://www.ingber.com/smni12_vectpot.pdf}.

\bibitem[Ingber(2012c)]{Ingber2012c}
L.~Ingber, ``Adaptive simulated annealing,'' in \emph{Stochastic global
  optimization and its applications with fuzzy adaptive simulated annealing},
  J.~H.A.~Oliveira, A.~Petraglia, L.~Ingber, M.~Machado, and M.~Petraglia,
  Eds.\hskip 1em plus 0.5em minus 0.4em\relax New York: Springer, 2012c, pp.
  33--61, {}Invited Paper. \url{https://www.ingber.com/asa11_options.pdf}.

\bibitem[Ingber(2013)]{Ingber2013}
L.~Ingber, ``Electroencephalographic {(EEG)} influence on {Ca2+} waves: Lecture
  plates,'' Lester Ingber Research, Ashland, OR, Tech. Rep. Report 2013:LEFI,
  2013, {}2nd World Neuroscience Online Conference 18 June 2013.
  \url{https://www.ingber.com/smni13_eeg_ca_lect.pdf}.

\bibitem[Ingber(2015)]{Ingber2015}
L.~Ingber, ``Calculating consciousness correlates at multiple scales of
  neocortical interactions,'' in \emph{Horizons in Neuroscience Research},
  A.~Costa and E.~Villalba, Eds.\hskip 1em plus 0.5em minus 0.4em\relax
  Hauppauge, NY: Nova, 2015, pp. 153--186, {}ISBN: 978-1-63482-632-7. Invited
  paper. \url{https://www.ingber.com/smni15_calc_conscious.pdf}.

\bibitem[Ingber(2016a)]{Ingber2016a}
L.~Ingber, ``Statistical mechanics of neocortical interactions: Large-scale {EEG}
  influences on molecular processes,'' \emph{Journal of Theoretical Biology},
  vol. 395, pp. 144--152, 2016a,
  {}\url{https://doi.org/10.1016/j.jtbi.2016.02.003}.

\bibitem[Ingber(2016b)]{Ingber2016b}
L.~Ingber, ``Path-integral quantum {PATHTREE} and {PATHINT} algorithms,''
  \emph{International Journal of Innovative Research in Information Security},
  vol.~3, no.~5, pp. 1--15, 2016b,
  {}\url{https://www.ingber.com/path16_quantum_path.pdf}.

\bibitem[Ingber(2017a)]{Ingber2017a}
L.~Ingber, ``Evolution of regenerative ca-ion wave-packet in neuronal-firing
  fields: Quantum path-integral with serial shocks,'' \emph{International
  Journal of Innovative Research in Information Security}, vol.~4, no.~2, pp.
  14--22, 2017a,
  {}\url{https://www.ingber.com/path17_quantum_pathint_shocks.pdf}.

\bibitem[Ingber(2017b)]{Ingber2017b}
L.~Ingber, ``Quantum path-integral {qPATHINT} algorithm,'' \emph{The Open
  Cybernetics Systemics Journal}, vol.~11, pp. 119--133, 2017b,
  {}\url{https://doi.org/10.2174/1874110X01711010119}.

\bibitem[Ingber(2017c)]{Ingber2017c}
L.~Ingber, ``Options on quantum money: Quantum path-integral with serial shocks,''
  \emph{International Journal of Innovative Research in Information Security},
  vol.~4, no.~2, pp. 7--13, 2017c,
  {}\url{https://www.ingber.com/path17_quantum_options_shocks.pdf}.

\bibitem[Ingber and Mondescu(2003)]{Ingber+Mondescu2003}
L.~Ingber and R.~Mondescu, ``Automated internet trading based on optimized
  physics models of markets,'' in \emph{Intelligent Internet-Based Information
  Processing Systems}, R.~Howlett, N.~Ichalkaranje, L.~Jain, and G.~Tonfoni,
  Eds.\hskip 1em plus 0.5em minus 0.4em\relax Singapore: World Scientific,
  2003, pp. 305--356, {}Invited paper.
  \url{https://www.ingber.com/markets03_automated.pdf}.

\bibitem[Ingber and Nunez(1995)]{Ingber+Nunez1995}
L.~Ingber and P.~Nunez, ``Statistical mechanics of neocortical interactions:
  High resolution path-integral calculation of short-term memory,''
  \emph{Physical Review E}, vol.~51, no.~5, pp. 5074--5083, 1995,
  {}\url{https://www.ingber.com/smni95_stm.pdf}.

\bibitem[Ingber and Nunez(2010)]{Ingber+Nunez2010}
L.~Ingber, ``Neocortical dynamics at multiple scales: {EEG} standing waves,
  statistical mechanics, and physical analogs,'' \emph{Mathematical
  Biosciences}, vol. 229, pp. 160--173, 2010,
  {}\url{https://www.ingber.com/smni10_multiple_scales.pdf}.

\bibitem[Ingber and Rosen(1992)]{Ingber+Rosen1992}
L.~Ingber and B.~Rosen, ``Genetic algorithms and very fast simulated
  reannealing: A comparison,'' \emph{Mathematical Computer Modelling}, vol.~16,
  no.~11, pp. 87--100, 1992, {}\url{https://www.ingber.com/asa92_saga.pdf}.

\bibitem[Ingber and Wilson(1999)]{Ingber+Wilson1999}
L.~Ingber and J.~Wilson, ``Volatility of volatility of financial markets,''
  \emph{Mathematical Computer Modelling}, vol.~29, no.~5, pp. 39--57, 1999,
  {}\url{https://www.ingber.com/markets99_vol.pdf}.

\bibitem[Ingber and Wilson(2000)]{Ingber+Wilson2000}
L.~Ingber, ``Statistical mechanics of financial markets: Exponential modifications
  to black-scholes,'' \emph{Mathematical Computer Modelling}, vol.~31, no. 8/9,
  pp. 167--192, 2000, {}\url{https://www.ingber.com/markets00_exp.pdf}.

\bibitem[Ingber et~al.(1991)Ingber, Fujio, and Wehner]{Ingber+Fujio+Wehner1991}
L.~Ingber, H.~Fujio, and M.~Wehner, ``Mathematical comparison of combat
  computer models to exercise data,'' \emph{Mathematical Computer Modelling},
  vol.~15, no.~1, pp. 65--90, 1991,
  {}\url{https://www.ingber.com/combat91_data.pdf}.

\bibitem[Ingber et~al.(1996)Ingber, Srinivasan, and
  Nunez]{Ingber+Srinivasan+Nunez1996}
L.~Ingber, R.~Srinivasan, and P.~Nunez, ``Path-integral evolution of chaos
  embedded in noise: Duffing neocortical analog,'' \emph{Mathematical Computer
  Modelling}, vol.~23, no.~3, pp. 43--53, 1996,
  {}\url{https://www.ingber.com/path96_duffing.pdf}.

\bibitem[Ingber et~al.(2001)Ingber, Chen, Mondescu, Muzzall, and
  Renedo]{Ingber+Chen+Mondescu+Muzzall+Renedo2001}
L.~Ingber, C.~Chen, R.~Mondescu, D.~Muzzall, and M.~Renedo, ``Probability tree
  algorithm for general diffusion processes,'' \emph{Physical Review E},
  vol.~64, no.~5, pp. 056\,702--056\,707, 2001,
  {}\url{https://www.ingber.com/path01_pathtree.pdf}.

\bibitem[Ingber et~al.(2014)Ingber, Pappalepore, and
  Stesiak]{Ingber+Pappalepore+Stesiak2014}
L.~Ingber, M.~Pappalepore, and R.~Stesiak, ``Electroencephalographic field
  influence on calcium momentum waves,'' \emph{Journal of Theoretical Biology},
  vol. 343, pp. 138--153, 2014,
  {}\url{https://doi.org/10.1016/j.jtbi.2013.11.002}.

\bibitem[Innocenti et~al.(2000)Innocenti, Parpura, and
  Haydon]{Innocenti+Parpura+Haydon2000}
B.~Innocenti, V.~Parpura, and P.~Haydon, ``Imaging extracellular waves of
  glutamate during calcium signaling in cultured astrocytes,'' \emph{Journal of
  Neuroscience}, vol.~20, no.~5, pp. 1800--1808, 2000.

\bibitem[Jensen(1987)]{Jensen1987}
A.~Jensen, ``Individual differences in the hick paradigm,'' in \emph{Speed of
  Information-Processing and Intelligence}, P.~Vernon, Ed.\hskip 1em plus 0.5em
  minus 0.4em\relax Norwood, NJ: Ablex, 1987, pp. 101--175.

\bibitem[Kozlowski et~al.(2015)Kozlowski, Caballero-Benitez, and
  Mekhov]{Kozlowski+Caballero-Benitez+Mekhov2015}
W.~Kozlowski, S.~Caballero-Benitez, and I.~Mekhov, ``Non-hermitian dynamics in
  the quantum zeno limit,'' U Oxford, Oxford, UK, Tech. Rep. arXiv:1510.04857
  [quant-ph], 2015, {}\url{https://arxiv.org/pdf/1510.04857.pdf}.

\bibitem[Langouche et~al.(1979)Langouche, Roekaerts, and
  Tirapegui]{Langouche+Roekaerts+Tirapegui1979}
F.~Langouche, D.~Roekaerts, and E.~Tirapegui, ``Discretization problems of
  functional integrals in phase space,'' \emph{Physical Review D}, vol.~20, pp.
  419--432, 1979.

\bibitem[Langouche et~al.(1982)Langouche, Roekaerts, and
  Tirapegui]{Langouche+Roekaerts+Tirapegui1982}
L.~Ingber, \emph{Functional Integration and Semiclassical Expansions}.\hskip 1em
  plus 0.5em minus 0.4em\relax Dordrecht, The Netherlands: Reidel, 1982.

\bibitem[Majhi and Ghosh(2018)]{Majhi+Ghosh2018}
S.~Majhi and D.~Ghosh, ``Alternating chimeras in networks of ephaptically
  coupled bursting neurons,'' \emph{Chaos}, vol.~28, no. 083113, 2018,
  {}\url{https://doi.org/10.1063/1.5022612}.

\bibitem[McKemmish et~al.(2009)McKemmish, Reimers, McKenzie, Mark, and
  Hush]{McKemmish+Reimers+McKenzie+Mark+Hush2009}
L.~McKemmish, J.~Reimers, R.~McKenzie, A.~Mark, and N.~Hush, ``Penrose-hameroff
  orchestrated objective-reduction proposal for human consciousness is not
  biologically feasible,'' \emph{Physical Review E}, vol.~80, no. 021912, pp.
  1--6, 2009, {}\url{https://doi.org/10.1103/PhysRevE.80.021912}.

\bibitem[Mountcastle et~al.(1981)Mountcastle, Andersen, and
  Motter]{Mountcastle+Andersen+Motter1981}
V.~Mountcastle, R.~Andersen, and B.~Motter, ``The influence of attentive
  fixation upon the excitability of the light-sensitive neurons of the
  posterior parietal cortex,'' \emph{Journal of Neuroscience}, vol.~1, pp.
  1218--1235, 1981.

\bibitem[Muller et~al.(2016)Muller, Gherardini, and
  Caruso]{Muller+Gherardini+Caruso2016}
M.~Muller, S.~Gherardini, and F.~Caruso, ``Quantum zeno dynamics through
  stochastic protocols,'' U Florence, Florence, Italy, Tech. Rep.
  arXiv:1607.08871 [quant-ph], 2016, {}\url{https://arxiv.org/pdf/1607.08871.pdf}.

\bibitem[Murakami and Okada(2006)]{Murakami+Okada2006}
S.~Murakami and Y.~Okada, ``Contributions of principal neocortical neurons to
  magnetoencephalography and electroencephalography signals,'' \emph{Journal of
  Physiology}, vol. 575, no.~3, pp. 925--936, 2006.

\bibitem[Nunez and Srinivasan(2006)]{Nunez+Srinivasan2006}
P.~Nunez and R.~Srinivasan, \emph{Electric Fields of the Brain: The
  Neurophysics of {EEG,} 2nd Ed}.\hskip 1em plus 0.5em minus 0.4em\relax
  London: Oxford University Press, 2006.

\bibitem[Nunez et~al.(2013)Nunez, Srinivasan, and
  Ingber]{Nunez+Srinivasan+Ingber2013}
P.~Nunez, R.~Srinivasan, and L.~Ingber, ``Theoretical and experimental
  electrophysiology in human neocortex: Multiscale correlates of conscious
  experience,'' in \emph{Multiscale Analysis and Nonlinear Dynamics: From genes
  to the brain}, M.~Pesenson, Ed.\hskip 1em plus 0.5em minus 0.4em\relax New
  York: Wiley, 2013, pp. 149--178,
  {}\url{https://doi.org/10.1002/9783527671632.ch06}.

\bibitem[Patil et~al.(2015)Patil, Chakram, and
  Vengalattore]{Patil+Chakram+Vengalattore2015}
Y.~Patil, S.~Chakram, and M.~Vengalattore, ``Measurement-induced localization
  of an ultracold lattice gas,'' \emph{Physical Review Letters}, vol. 115, no.
  140402, pp. 1--5, 2015, {}\url{https://doi.org/10.1103/PhysRevLett.115.140402}.

\bibitem[Preskill(2015)]{Preskill2015}
J.~Preskill, ``Quantum mechanics,'' Caltech, Pasadena, CA, Tech. Rep. Lecture
  Notes, 2015, {}http://www.theory.caltech.edu/people/preskill/ph219/.

\bibitem[Reyes and Parpura(2009)]{Reyes+Parpura2009}
R.~Reyes and V.~Parpura, ``The trinity of {Ca2+} sources for the exocytotic
  glutamate release from astrocytes,'' \emph{Neurochemistry International},
  vol.~55, no.~3, pp. 1--14, 2009.

\bibitem[Ross(2012)]{Ross2012}
W.~Ross, ``Understanding calcium waves and sparks in central neurons,''
  \emph{Nature Reviews Neuroscience}, vol.~13, pp. 157--168, 2012.

\bibitem[Salazar et~al.(2012)Salazar, Dotson, Bressler, and
  Gray]{Salazar+Dotson+Bressler+Gray2012}
R.~Salazar, N.~Dotson, S.~Bressler, and C.~Gray, ``Content-specific
  fronto-parietal synchronization during visual working memory,''
  \emph{Science}, vol. 338, no. 6110, pp. 1097--1100, 2012,
  {}\url{https://doi.org/10.1126/science.1224000}.

\bibitem[Scemes and Giaume(2006)]{Scemes+Giaume2006}
E.~Scemes and C.~Giaume, ``Astrocyte calcium waves: What they are and what they
  do,'' \emph{Glia}, vol.~54, no.~7, pp. 716--725, 2006,
  {}\url{https://doi.org/10.1002/glia.20374}.

\bibitem[Schulman(1981)]{Schulman1981}
L.~Schulman, \emph{Techniques and Applications of Path Integration}.\hskip 1em
  plus 0.5em minus 0.4em\relax New York: J. Wiley Sons, 1981.

\bibitem[Schulten(1999)]{Schulten1999}
K.~Schulten, ``Quantum mechanics,'' U. Illinois, Urbana, IL, Tech. Rep. PHYS480
  Lecture Notes, Chapter 2, 1999,
  {}\url{http://www.ks.uiuc.edu/Services/Class/PHYS480/}.

\bibitem[Volterra et~al.(2014)Volterra, Liaudet, and
  Savtchouk]{Volterra+Liaudet+Savtchouk2014}
A.~Volterra, N.~Liaudet, and I.~Savtchouk, ``Astrocyte {Ca2+} signalling: an
  unexpected complexity,'' \emph{Nature Reviews Neuroscience}, vol.~15, pp.
  327--335, 2014.

\bibitem[Wehner and Wolfer(1983a)]{Wehner+Wolfer1983a}
M.~Wehner and W.~Wolfer, ``Numerical evaluation of path-integral solutions to
  fokker-planck equations. {I},'' \emph{Physical Review A}, vol.~27, pp.
  2663--2670, 1983a.

\bibitem[Wehner and Wolfer(1983b)]{Wehner+Wolfer1983b}
M.~Wehner and W.~Wolfer, ``Numerical evaluation of path-integral solutions to fokker-planck
  equations. {II.} restricted stochastic processes,'' \emph{Physical Review A},
  vol.~28, pp. 3003--3011, 1983b.

\bibitem[Wehner and Wolfer(1987)]{Wehner+Wolfer1987}
M.~Wehner and W.~Wolfer, ``Numerical evaluation of path integral solutions to fokker-planck
  equations. {III.} time and functionally dependent coefficients,''
  \emph{Physical Review A}, vol.~35, pp. 1795--1801, 1987.

\bibitem[Wu et~al.(2012)Wu, Wang, and Yi]{Wu+Wang+Yi2012}
S.~Wu, L.~Wang, and X.~Yi, ``Time-dependent decoherence-free subspace,''
  \emph{Journal of Physics A}, vol. 405305, pp. 1--11, 2012.

\bibitem[Zhang and Simon(1985)]{Zhang+Simon1985}
G.~Zhang and H.~Simon, ``{STM} capacity for chinese words and idioms: Chunking
  and acoustical loop hypotheses,'' \emph{Memory Cognition}, vol.~13, pp.
  193--201, 1985.

\bibitem[Zhang et~al.(2014)Zhang, Ai, Li, Xu, and Sun]{Zhang+Ai+Li+Xu+Sun2014}
P.~Zhang, Q.~Ai, Y.~Li, D.~Xu, and C.~Sun, ``Dynamics of quantum zeno and
  anti-zeno effects in an open system,'' \emph{Science China Physics, Mechanics
  Astronomy}, vol.~57, no.~2, pp. 194--207, 2014,
  {}\url{https://doi.org/10.1007/s11433-013-5377-x}.

\end{thebibliography}
\end{document}